\input harvmac
\input epsf
\baselineskip=11pt
\overfullrule=0pt
\def\nl{\hfil\break}

\def\half{{1\over2}}
\def\c#1{{\cal #1}}
\def\IR{\relax{\rm I\kern-.18em R}}

\def\corru#1#2{\langle #2\rangle_{{}_{#1}}}
\def\prodone#1{\prod_{j=1}^{#1}}
\def\del{\partial}
\def\Re{{\rm Re}}

\def\pint{{\int \!\!\!\!\!\! -}}
\def\al#1{a_{#1}}
\def\em#1{\it #1}
\def\qq#1{q_{#1}}
\def\section#1{\newsec{#1}}

\def\ijp{{ \sl Int. J. of Mod. Phys. }}

\def\npb{{ \sl Nucl. Phys. }}

\def\plb{{ \sl Phys. Lett. }}

\font\cmss=cmss10 \font\cmsss=cmss10 at 7pt
\def\IZ{\relax\ifmmode\mathchoice
  {\hbox{\cmss Z\kern-.4em Z}}{\hbox{\cmss Z\kern-.4em Z}}
  {\lower.9pt\hbox{\cmsss Z\kern-.4em Z}}
  {\lower1.2pt\hbox{\cmsss Z\kern-.4em Z}}\else{\cmss Z\kern-.4em Z}\fi}
\Title{UCLA/96/TEP/27}{\titlefont 
  Non--Critical Strings At High Energy }
\centerline{Kenichiro Aoki\footnote{$^1$}
{Internet electronic mail address:{\tt~ken@hc.keio.ac.jp}}
and Eric D'Hoker\footnote{$^2$}
{Internet electronic mail address:{\tt~dhoker@physics.ucla.edu}}}
\vskip.8cm
\centerline{\it ${}^1$Department of Physics, Hiyoshi Campus} 
 \centerline{\it Keio University}
  \centerline{\it Hiyoshi, Kouhoku-ku}
  \centerline{\it Yokohama { 223},   JAPAN}
\vskip.3in
\centerline{\it ${}^2$ Theory Division, CERN}
\centerline{\it 1211 Geneva 23, Switzerland}
\centerline{\it and}
\centerline{\it Department of Physics}
\centerline{\it    University of California Los Angeles}
\centerline{{\it   Los Angeles, California}
   90024{\rm --}1547\it,  USA}
\vskip .4cm
\centerline{{\bf Abstract}}
\bigskip
We consider scattering amplitudes in non-critical string theory
of $N$ external states in the limit where the energy of all
external states is large compared to the string tension. 
We argue that the amplitudes are naturally complex analytic in 
the matter central charge $c$ and we propose to define the
amplitudes for arbitrary value of $c$ by analytic continuation.
We show that the high energy limit is dominated by a saddle 
point that can be mapped onto an equilibrium electro-static 
energy configuration of an assembly of $N$ pointlike (Minkowskian) 
charges, together with a density of charges arising from the 
Liouville field.  We argue that the Liouville charges accumulate 
on segments of curves, and produce quadratic branch cuts on the 
worldsheet. The electro-statics problem is solved for string 
tree level in terms of hyper-elliptic integrals and is given 
explicitly for 3- and 4-point functions. We show that the high
energy limit should behave in a string-like fashion with
exponential dependence on the energy scale for generic
values of $c$.
\Date{}
\vfill\eject
\section{Introduction}
Ever since string theory was reformulated 
in terms of a summation over random surfaces,
\ref\POLYAKOV{
A.M. Polyakov, Phys. Lett. {\bf B103} (1981) 207, 211}, 
there has been a renewed and persistent interest
in the construction of consistent string theories away
from the critical dimensions (26 for the bosonic string, 10 for
superstrings),
\ref\STRINGA{
T. Curtright and C. Thorn, Phys. Rev. Lett. {\bf 48} (1982) 1309;
E. D'Hoker and R. Jackiw, Phys. Rev. {\bf D26} (1982) 3517;
Phys. Rev. Lett. {\bf 50} (1983) 1719;
J.-L. Gervais and A. Neveu, Nucl. Phys. {\bf B199} (1982) 59; 
Nucl. Phys. {\bf B209} (1982) 125}, 
\ref\GERNEVA{
J.-L. Gervais and A. Neveu, Phys. Lett. {\bf B123} (1983) 86; 
Nucl. Phys. {\bf B224} (1983) 329;
J.-L. Gervais and A. Neveu, Nucl. Phys. {\bf B238 } (1984) 125; 
Comm. Math. Phys. {\bf 100} (1985) 15
},
\ref\GERNEVB{
J.-L. Gervais and A. Neveu, Phys. Lett. {\bf B151} (1985) 271; 
Nucl. Phys. {\bf B264} (1986) 557;
A. Bilal and J.-L. Gervais, Phys. Lett. {\bf B187 } (1987) 39; 
Comm. Math. Phys. {\bf 100} (1985) 15
},
\ref\CT{
T. Curtright and C. Thorn, Phys. Lett. {\bf B118} (1982) 115;
E. Braaten, T. Curtright and C. Thorn, Ann. Phys. (N.Y.) 
{\bf 147} (1983) 365;
E. Braaten, T. Curtright, G. Ghandour and C. Thorn, 
Phys. Rev. Lett. {\bf 51} (1983) 19; Ann. Phys. (N.Y.) {\bf 153}
(1984) 147},
\ref\STRINGB{
A. Belavin, A.M. Polyakov and A.B. Zamolodchikov, Nucl. Phys. 
{\bf B241} (1984) 333;
A.M. Polyakov, Mod. Phys. Lett. {\bf A2} (1987) 893;
V. Knizhnik, A.M. Polyakov and A.B. Zamolodchikov, Mod. Phys. 
Lett. {\bf A3} (1988) 819},
\ref\LATTICE{
F. David, Nucl. Phys. {\bf B257} (1985) 45, 543;
J. Ambjorn, B. Durhuus and J. Fr\"ohlich, Nucl. Phys. 
{\bf B257} (1985) 433;
V.A. Kazakov, I.K. Kostov and A.A. Migdal, Phys. Lett. 
{\bf B157} (1985) 295;
D. Boulatov, V.A. Kazakov, I.K. Kostov and A.A. Migdal, Phys.
Lett. {\bf B174} (1986) 87; Nucl. Phys. {\bf B275} (1986)
}
\ref\STRINGC{
F. David and E. Guitter, Euro. Phys. Lett. {\bf 3} (1987) 1169;
F. David, Mod. Phys. Lett. {\bf A3} (1988) 1651;
J. Distler and H. Kawai, Nucl. Phys. {\bf B321} (1988) 509
},
\ref\MATRIX{
M. Douglas and S. Shenker, Nucl. Phys. {\bf B335} (1990) 509;
E. Br\'ezin and V. Kazakov, Phys. Lett. {\bf B236} (1990) 144;
D.J. Gross and  A. Migdal, Phys. Rev. Lett. {\bf 64} (1990) 127; 
Nucl. Phys. {\bf B340} (1990) 333
},
\ref\WITTEN{
E. Witten, Nucl. Phys. {\bf B340} (1990) 281;
R. Dijkgraaf and E. Witten, Nucl. Phys. {\bf B342} (1990) 486;
R. Dijkgraaf, E. Verlinde and H. Verlinde, Nucl. Phys. {\bf B352} (1991) 59
}. 
Reviews on and extensive references to work on non-critical 
string theory may be found in
\ref\REV{
N. Seiberg, {\it Notes on Quantum Liouville Theory and Quantum 
Gravity}, Progr. Theor. Phys. Suppl. {\bf 102} (1990) 319;
F. David, {\it Non-perturbative Effects in 2D Gravity 
and Matrix Models},
in Proceedings Cargese Workshop 1990, edited by O. Alvarez,
V. Kazakov and P. Windey, NATO ASI Series B262;
J. Polchinski, {\it Remarks on Liouville Field Theory}, 
Strings 90
Conference, College Station, Texas, Published in College Station 
Workshop, World Scientific Publ. (1991);
D.J. Gross, {\it The $c=1$ Matrix Models}, in Proceedings 
to Jerusalem Winter School, 1990;
E. D'Hoker, {\it Lecture Notes on 2-D Quantum Gravity 
and Liouville Theory}, in ``Particle Physics VI-th 
Jorge Andre Swieca Summer School'',
ed. O.J.P. Eboli, M. Gomes and A. Santoro, World 
Scientific Publishers, (1992);
P. Ginsparg and G. Moore, {\it Lectures on 2D Gravity and 2D String 
Theory}, TASI Lecture Notes, TASI summer School 1992, 
Published in Boulder
TASI 1992;
F. David, {\it Symplicial Quantum Gravity and Random Lattices}, 
hep-th/9303127;
"2-D Gravity in Non-Critical Strings", Discrete and Continuum Approaches,
E. Abdalla, M.C.B. Abdalla, D. Dalmazi, A. Zadra,
Springer-Verlag, 
Lecture Notes in Physics, {\bf 20}, 1994
} and
\ref\DGZ{
P. Di Francesco, P. Ginsparg and J. Zinn-Justin, Physics Reports 
{\bf 254} (1995) 1, and references therein.}.
Several motivations are driving these studies, of which we shall 
just mention the most immediate ones. There is the uncovering of
more general or new consistent string theories : for example a six 
dimensional tensionless string theory has recently been considered
\ref\TENSIONLESS{
J.H. Schwarz, hep-th/9604171;
Berkooz, R. Leigh, J. Polchinski, N. Seiberg and E. Witten, 
hep-th/9605184}.
There is the improvement of our understanding -- perhaps at the
non-perturbative level -- of critical string theory via models with
fewer physical degrees of freedom. There is the reinterpretation of
the worldsheet properties of the string in terms of two-dimensional
quantum gravity, providing simple examples of quantized gravity
\STRINGB. There is the possibility of mapping the scaling behavior
of  two-dimensional statistical mechanics models on random lattices
onto 2D quantum gravity
\LATTICE, \MATRIX. There is the proposed reformulation of the critical
behavior of the three-dimensional  Ising model in terms of fermionic
strings 
\ref\ISINGA{
A.M. Polyakov, Phys. Lett. {\bf B8} (1979) 247;
S. Samuel, J. Math. Phys. {\bf 21} (1980) 2806, 2815, 2820;
E. Fradkin, M. Srednicki and L. Susskind, Phys. Rev. {\bf D21}
 (1980) 2885;
A. Casher, D. Foerster and P. Windey, Nucl. Phys. {\bf B251} 
(1985) 29}%
\ref\ISINGB{
Vl.S. Dotsenko and A.M. Polyakov, ``Fermion Representations for 
the 2-D and 3-D Ising Models'', in {\it Kyoto 1986, Proceedings, 
Conformal Field Theory and Solvable Models}, p. 171;
Vl.S. Dotsenko, Nucl. Phys. {\bf B285} (1987) 45;
A.R. Kavalov and A.G. Sedrakyan, Nucl. Phys. {\bf B285} (1987) 264;
Vl.S. Dotsenko, M. Picco, P. Windey, G. Harris, E. Martinec and 
E. Marinari, Nucl. Phys. {\bf B448} (1995) 577
}%
\ref\ISINGC{
J.-L. Gervais and A. Neveu, Nucl. Phys. {\bf B257} (1985) 59; 
A. Bilal and J.-L. Gervais, Nucl. Phys. {\bf B295} (1988) 277
}.
There are proposals to view non-critical string theory
as a conformally invariant off-shell realization of critical
string theory
\ref\PM{
E. D'Hoker, unpublished (1987);
R.C. Meyers and V. Periwal, Phys. Rev. Lett. {\bf 70} (1993) 2841;
E. Kiritsis and C. Kounnas, Nucl. Phys. {\bf B442} (1995) 472
},
and as a critical string theory in a flat space-time background 
metric with a dilaton field that grows linearly with time
\ref\KK{
R. Myers, \plb{\bf B199} (1987) 371; 
I. Antoniadis, C. Bachas, J. Ellis, D. Nanopoulos, \plb{\bf
  B211} (1988) 393; \npb{\bf B328} (1989) 115; 
J. Polchinski, \npb{\bf B324} (1989) 123, and references
}.

Our understanding of non-critical string theory to date is very
advanced for bosonic models with rational matter 
central charge $c<1$. The mapping between 
discretized random surfaces and random matrices, combined with
the double scaling limit, produces exact results for correlation
functions, to all orders of perturbation theory \LATTICE, 
\MATRIX, see \DGZ\ for a review.  
Surprisingly,
it has turned out to be very difficult to reproduce, with
the help of the Liouville model of Polyakov's proposal, even the
simplest results obtained via matrix models. Also, a
direct reformulation of fermionic non-critical strings in terms
of matrix models seems problematic, if not impossible
\ref\SUPERMATRIX{
  See for instance, 
  L. Alvarez-Gaume, H. Itoyama, J.L. Manes, A. Zadra,
  \ijp{\bf A7} (1992) 5337;
  R. Brustein, M. Faux, B.A. Ovrut \npb{\bf B421}(1994) 293
  
}. 
Finally, it has proven to be quite difficult to cross the barrier at
$c=1$ and analyze the region of perhaps most pronounced physical
interest $c>1$, for either the bosonic or fermionic strings.

One of the most basic obstacles to reaching beyond the $c=1$
barrier (within the Liouville field theory approach), is the
appearance of conformal primary fields with complex weights, and
thus of string states with complex masses. In a series of 
ingenious papers \GERNEVA\ and \GERNEVB, it was proposed that
the string spectrum may be restricted to a subset of ``physical
states", that have real conformal weights only. This restriction 
appears to be possible only provided 
space-time dimension assumes certain special values : 1, 7, 13 and 19
for the bosonic string, and 1, 3, 5 and 7 for the fermionic string.
(The truncation of the spectrum in \GERNEVA and \GERNEVB is 
analogous to the truncation of the Kac table of conformal primary 
fields used in \STRINGB\ for rational conformal field theories with 
$c<1$.)
\foot{See also
\ref\KS{
D. Kutasov and N. Seiberg, Phys. Lett. {\bf 251} (1990) 67},
where a truncation, analogous to the GSO projection, was 
proposed for the non-critical fermionic string.
}

A truncation of the spectrum at the free string level will be 
consistent at the interacting level, only if interactions 
between physical states produce only physical states. 
The simplest direct check would be on the factorization of the four 
point function at tree level; unfortunately, this amplitude 
is not available. Instead, it was verified in
a series of papers 
\ref\QGROUP{
J.L. Gervais, Comm. Math. Phys. {\bf 130} (1990) 257;
{\bf 138} (1991) 301; J. Mod. Phys. {\bf B6} (1992);
J.-L. Gervais, Nucl. Phys. {\bf B391} (1993) 287;
E. Cremmer, J.-L. Gervais and J.-F. Roussel, Comm. Math. Phys. 
{\bf 161} (1994) 597;
E. Cremmer, J.-L. Gervais and J.-F. Roussel, Nucl. Phys. 
{\bf B413} (1994) 244, 433;
J.-L. Gervais and J.-F. Roussel, Phys. Lett. {\bf B338} (1994) 437;
J.-L. Gervais and J.-F. Roussel, Nucl. Phys. {\bf B426} (1994) 140
}, and 
\ref\SCHNITTGER{
J.-L. Gervais and J. Schnittger, Phys. Lett. {\bf B315} (1993) 258;
Nucl. Phys. {\bf B413} (1994) 433; E. Cremmer, J.-L. Gervais and J.
Schnittger, hep-th/9503198; hep-th/9604131; J.-L. Gervais,
hep-th/9606151
},
that the algebra of operators corresponding to ``physical states"
closes under operator product
expansion, again provided the dimension of space-time belongs
to the list given above. This somewhat less direct check on
the truncation of string states provides strong evidence that
consistent string theories indeed exist in these special dimensions.
Clearly, however, it would be very valuable to have access to
the four point function for a more direct check of factorization.

The primary goal of this paper is to develop calculational methods
that allow us to evaluate scattering amplitudes in non-critical 
string theory. We shall show that this can indeed be achieved, in the
limit where the energies of the incoming and outgoing strings
is large.

We propose to define non-critical string
amplitudes in the region $1<c<25$ by analytic continuation in
$c$ throughout the complex $c$ plane,
\foot{Analytic continuations in the
  central charge were used in 
  \ref\AD{K. Aoki and E. D'Hoker, Mod. Phys. Lett. {\bf A3} (1992) 235 }
  as technical tools to prove the validity of the continuation
  procedure of 
  \ref\GL{M. Goulian and M. Li, Phys. Rev. Lett. {\bf 66} (1991) 2051}.  
  Here, we go one step further, and
  take the analytic continuation as a definition of the
  amplitudes.}
starting from the line $c<1$. As was already shown in \AD, 
the integral representation of the non-critical scattering 
amplitudes in the Liouville formulation, is complex 
analytic in the central charge,
and thus naturally lends itself to such a definition. Also,
this type of analytic continuation is very similar to that
required to defining scattering amplitudes for all values of
external momenta, already in the critical string. (See {\it e.g.}
\ref\EDDP{
E. D'Hoker and D.H. Phong, Phys. Rev. Lett. {\bf 70} (1993) 3692;
Nucl. Phys. {\bf B440} (1995) 24}.)

The definition of non-critical string amplitudes by analytic
continuation in the central charge that we use as a starting
point is a priori different from the one used in \QGROUP\ and
\SCHNITTGER. There, substantial modifications occur in the 
structure of the Liouville field dynamics as one crosses from
the weak coupling phase (for $c<1$) into the strong coupling
phase (for $c>1)$. In particular, the two chiralities of the Liouville
field become uncoupled, a new cosmological constant appears and 
the vertex operators are modified. Whether the definition of
the amplitudes by analytic continuation in the central charge
$c$ takes us to the non-critical strings of \GERNEVA\ and \GERNEVB\
is an open and exciting question.

The exact evaluation of the scattering amplitudes, even to tree level,
for general $c$ and general external 
momenta, would require that we can carry out a set of multiple 
integrals that are more general than those available from 
\ref\DF{
Vl.S. Dotsenko and V.A. Fateev, Nucl. Phys. {\bf B240} (1984) 312; 
{\bf B251} (1985) 691
}, 
or from
\ref\DIKUT{
P. Di Francesco and D. Kutasov, Phys. Lett. {\bf B261} (1991) 385;
P. Di Francesco and D. Kutasov, Nucl. Phys. {\bf B342} (1990) 475;
Nucl. Phys. {\bf B375} (1992) 119
}. Evaluating these integrals remains an open problem.

In this paper, we propose an evaluation of non-critical string
amplitudes for any complex $c$, in the limit where the energies
of incoming and outgoing string states are all large compared to
the (square root of the) string tension.%
\foot{
  The high energy limit of non-critical string theories 
  with $c=1$ was considered in
  \ref\JLY{
    A. Jevicki, M. Li and T. Yoneya, Nucl. Phys. {\bf B448} (1995) 277
    }.
  }
We shall show that the Liouville approach lends itself naturally 
to taking the high energy limit, where the integral 
representations for the amplitudes become tractable, for any
complex value of $c$. To string tree level, we succeed in 
producing explicit formulas for the limit in terms of 
hyper-elliptic integrals. We shall not, at this stage, 
perform any truncation on the spectrum of states in the 
non-critical string theory. Thus, our results are applicable
to non-critical string theories in general, including 
those in which the Liouville field is reinterpreted as
an extra dimension of space-time, as in \KK.

For string theory in the critical dimension, the high energy
limit of scattering amplitudes is dominated by a saddle point in
the positions of the vertex operators for external string
states, as well as in the moduli of the surface.  This problem
is equivalent to finding the equilibrium configuration of an
array of electro-static Minkowskian charges (attached to the
vertex operators) on a surface of variable shape. In a series 
of beautiful papers 
\ref\GM{
D.J. Gross and P. Mende, Phys. Lett. {\bf B197} (1987) 129,
\npb{\bf B303 } (1988) 407;
D.J. Gross, Phys. Rev. Lett. {\bf 60} (1988) 1229;
D. Amati, M. Ciafaloni, G. Veneziano, \ijp{\bf A3} (1988) 1615, 
\plb{\bf 197B} (1987) 81
}, 
it was shown how the
saddle point can be constructed by symmetry arguments, for the
four point function, to any order in perturbation theory.

For non-critical string theory, the high energy limit is still
dominated by a saddle point, which is equivalent to the equilibrium
configuration of an array of (complex) charges on a surface of
variable shape. In addition to the charges from the external
vertex operators, we now also have charges from the Liouville
exponential operator. In fact, the number of Liouville charges
on the surface increases linearly with energy and, in the limit
of large energy, accumulate onto a continuous charge density. We
shall show that this Liouville charge density consists of line
segments, producing quadratic branch cuts on the worldsheet.

We shall solve explicitly the equivalent electro-statics problem
for a worldsheet with the topology of a sphere (tree level) in
terms of hyper-elliptic functions, and use it to deduce the high
energy limit of tree level scattering amplitudes.  The solution
in this limit is valid for any complex value of the matter central charge
$c$, and we use analytic continuation to define the non-critical
string amplitudes throughout the complex $c$ plane. For higher
genus topologies, the solution involves quadratic branch cuts of
higher genus surfaces, but we shall postpone a full derivation
of this case to a later publication.

The main result is that, at least for generic values of the 
matter central charge $c$, the non-critical amplitudes behave
in a string like fashion, with exponential dependence on the
energy scale, in the limit of high energy. While it is logically
possible that this generic exponential behavior could be
absent (and replaced by power-like behavior)
at isolated points in the complex $c$ plane, we believe
that this is unlikely to occur in the region $1<c<25$.
It is thus unlikely that the non-critical string theories
in this region ever become ``quantum field theories''.

The remainder of this paper is organized as follows. 
In Sect. 2, we establish the equivalence between the high 
energy limit saddle point and the electro-static 
equilibrium configuration of an array of charges, 
including Liouville charges, on a Riemann surface with 
variable moduli. 
In Sect. 3, we solve the electro-statics problem at
string tree level (i.e. on the complex plane), find the
configuration of charges and determine their electro-static
energy in terms of hyper-elliptic integrals. 
In Sect. 4, we work out the cases of the string tree-level 
3- and 4-point functions in detail, and use the symmetric
scattering amplitude as a simple explicit example. 
In Sect. 5, we present a brief discussion of open problems, 
in particular of the higher loop case, of the possibility of 
power law behavior and of some practical applications.
Some useful formulas are derived in Appendix A.
\section{High Energy Limit and Equivalent Electro-statics}
We begin by reviewing some basic results in the Liouville field
theory formulation of bosonic non-critical string theory. The
starting point is a ``matter'' conformal field theory,
describing Poincar\'e invariant string dynamics in a
$d$-dimensional space-time. In addition to the string
coordinates $x^\mu (z)$ with $\mu =1,\cdots d$, there may be
further ``internal degrees of freedom'', collectively denoted by
$\psi (z)$ in what follows. The worldsheet metric is denoted by 
$g_{mn}$ and the associated Laplacian on scalar functions by 
$\Delta _g$. The action for the string
coordinate $x$ is given by free field theory%
\foot{
The string tension $T$ can naturally be absorbed into $x$, and 
will be set equal to 1 in the remainder.}
\eqn\freex{
  S_M={T \over 4\pi} \int d^2z\sqrt{ g} ~  x ^\mu 
  \Delta _{ g} x _\mu
  }
The conformal primary fields are vertex operators of the type 
\eqn\confvertex{
                V_\delta = {\c P}_\Delta (\del x^\mu, \psi) 
                e^{ik \cdot x}
\qquad \qquad \delta= \Delta + \half k^2
} 
where ${\c P}_\Delta $ depends on $\psi$ and derivatives of $x^\mu$
only. The conformal dimensions of $V_\delta $ and ${\c P}_\Delta$ are
$(\delta, \delta)$ and $(\Delta, \Delta)$ respectively.  Without
loss of generality, we may consider vertex operators associated
to external string states of definite mass and spin; the factor
${\c P}_\Delta$ then grows with momenta (and energy) no faster than
polynomially.

The above conformal field theory is coupled to a quantized
worldsheet metric $g$, which, in conformal gauge, decomposes
into the Liouville field $\phi (z)$ and a fiducial metric $\hat
g(m_j)$ that only depends on the moduli $m_j$ of the Riemann
surface $\Sigma$, with $g=\hat g \exp\{2\phi\}$.  The action for
the Liouville field is
\eqn\laction{
  S_{L} ={1\over4\pi} \int\sqrt{\hat g} \left[
  \half \phi \Delta _{\hat g} \phi -\kappa  R_{\hat g} \phi
  +{ \mu }  e^{ \alpha\phi}        \right]
}
Here $R_{\hat g}$ is the Gaussian curvature of the metric 
$\hat g$, and the
coupling constants $\kappa$ and $\alpha$ are given in terms 
of the matter
central charge $c$ as follows
\eqn\kdef{
  3\kappa ^2 = 25-c \qquad \qquad  \qquad \qquad
  \alpha ^2 + \kappa \alpha +2 =0
}
Each conformal primary field $V_\delta$ may be coupled to the
worldsheet metric in a diffeomorphism invariant way. In
conformal gauge this is achieved by multiplying $V_\delta$ by a
Liouville exponential $\exp\{\beta(\delta)\phi\}$, and adjusting
$\beta (\delta)$ in such a way that the resulting operator has
conformal weight $(1,1)$. This gravitationally dressed operator
may be integrated in a diffeomorphism invariant way and we obtain
\eqn\vertexop{
  \c V_\delta \equiv \int d^2z V_\delta (z)
  e^{\beta\phi(z)},\qquad 
  \beta(\delta ) ={-\sqrt{25-c}+\sqrt{1-c+24\delta
      }\over2\sqrt3} 
}
The analytic continuation in $c$ and the external momentum will
dictate which branches of the square roots should be chosen.
When $\delta =0$, the operator $\c V _0$ is just the Liouville
exponential interaction, present in the Liouville
action.%
\foot{For special values of $c$, it is possible to have
  primary fields other than just exponentials of the Liouville
  field, so that combinations other than \vertexop\ may
  occur. This happens at $c=25$ and $c=-2$ for example.  Thus,
  the operators we are considering in \vertexop\ may not be the
  most general physical operators possible in non--critical
  string theory. Also, we do not know in general which operators
  correspond to the complete set of physical states in a
  particular theory.  A careful analysis of the tree level four
  point function which we compute below should shed light on
  these questions.  }

Correlation functions of the operators $\c V_\delta $ are
obtained in standard fashion by combining matter and Liouville
correlation functions%
\foot{To simplify notation, we denote $V_{\delta _i}$, 
${\cal V}_{\delta _i}$, and $\beta (\delta _i)$ by $V_i$, 
${\cal V}_i$ and $\beta _i$ respectively.}
\eqn\eqcorr{\corru{}{\prod_{i=1}^N\c V_i}=
  \sum_{h=0}^\infty\int_{\c M_h} \!\!\! dm Z_{gh}(m)
  \int_{\Sigma _{h}}
  \prod_{i}d^2z_i 
  \corru{L}{\prod_{i=1}^N e^{\beta_i\phi(z_i)}}
  \corru{M}{\prod_{i=1} ^N  V_i(z_i)}
}
Here, $\c M_h$ is the moduli space of compact Riemann surfaces
$\Sigma _{h}$ with $h$ handles, $dm$ stands for the measure on
moduli space, and $Z_{gh}(m)$ consists of Fadeev-Popov ghost
determinants, including their zero mode normalization
factors (see e.g. 
\ref\DHP{
E. D'Hoker and D.H. Phong, Rev. Mod. Phys {\bf 60} (1988) 917
}). 
The conformal field theory correlation functions for
the matter integrals and Liouville integrals are respectively
defined by
\eqn\corrdef{\eqalign{
    \corru{M}{\prod_{i=1} ^N  V_i(z_i)}
  &\equiv \int D_{\hat g}x \int D_{\hat g}\psi ~e^{-S_{M} -S_{\psi}} 
  \prod_{i}  V_i(z_i)
  \cr
\corru{L}{\prod_{i=1}^N e^{\beta_i\phi(z_i)}}
   & \equiv\int D_{\hat g}\phi ~e^{-S_{L}}
  \prod_{i} e^{\beta_i\phi(z_i)}\cr
  }
}
$S_{\psi}$ denotes the action for the additional internal
degrees of freedom, but we shall not need it here.  The
functional measures $D_{\hat g}x$ $D_{\hat g}\psi$ and $D_{\hat
  g}\phi$ are built from the ${\bf L}^2$ norms on the functions
with respect to the fiducial metric $\hat g$ (see
\STRINGC, 
\ref\DHK{N. Mavromatos and J. Miramontes, Mod. Phys. Lett. 
{\bf A4} (1989) 1849; 
E. D'Hoker and P. Kurzepa, Mod. Phys. Lett. {\bf A5} (1990) 1411; 
E. D'Hoker, Mod. Phys. Lett. {\bf A6} (1991) 795}).
\subsec{Correlation functions as multiple integrals}
In what follows, we shall compute Liouville and matter
correlation functions in the high energy limit, which will allow
us to evaluate correlation functions in non--critical string
theory in the same approximation.

To evaluate the Liouville correlation functions, we follow the
procedure of \GL, \AD
\ref\DK{
Vl.S. Dotsenko, Mod. Phys. Lett. {\bf A6} (1991) 3601;
Y. Kitazawa, Phys. Lett. {\bf B265} (1991) 262
} (see also
\ref\DO{
J. Teschner, Nucl. Phys. {\bf B413} (1994) 277;
Phys. Lett. {\bf B363} (1995) 65;
H. Dorn and H.-J. Otto, Nucl. Phys. {\bf B429} (1994) 375
}).
We split the Liouville field $\phi$ as follows $\phi = \phi _0 +
\varphi$, where $\phi _0$ is constant on the worldsheet and
$\varphi$ is orthogonal to constants. The integration splits
accordingly, and the integral over $\phi _0$ may be carried out
explicitly, as follows
\eqn\split{
  \int D_{\hat g}\phi\,e^{-S_L}\prodone N e^{\beta_j\phi(z_j)} =
  { \Gamma(-s) \mu ^s \over \alpha (4\pi)^s}
  \int D_{\hat g}\varphi \ e^{-S'_L}\left(\int\!\!\sqrt{\hat g}\
    e^{\alpha\varphi}\right)^s
  \prodone N e^{\beta_j\varphi(z_j)}
  }
Here $D_{\!\hat g}\varphi$ denotes the integration over the
field $\varphi$, which is orthogonal to constants, by
definition. The new Liouville action $S_L'$ is a free action
now, given by
\eqn\eqnonzeromodes{
  S'_L = {1\over4\pi} \int\!d^2z\sqrt{\hat g} \left[
  \half \varphi \Delta _{\hat g} \varphi -\kappa  R_{\hat g} \varphi
\right]
}
The variable $s$ is a scaling dimension, given in terms of
$\alpha$, $\kappa$, the genus $h$ of the surface and the
energies $\beta _j$ as follows 
\eqn\eqsdef{
  \alpha s=-{\kappa}(1-h)-\sum_{j=1}^N\beta_j
  }
In general, $s$ does not have to be integer, or does not even
have to be rational. The prescription of \GL\ is to
proceed and carry out the functional integration over $\varphi$
as if $s$ were an integer, and then later on continue in $s$. We
are confident that this procedure is reliable in view of the
semi-classical analysis carried out in \AD. 

The integration over $\varphi$ is completely analogous to the
Coulomb gas problem, but here with generalized complex charges.
\eqn\lcorr{\eqalign{
  \corru{L}{\prod_{j=1}^N e^{\beta_j\phi(z_j)}}
  & =
  {\Gamma(-s) \mu ^s\over\alpha (4\pi)^s} Z_s(m)^{-1/2}\int 
  \prod_{p=1}^sd^2w_a \exp\biggl \{
  \sum_{i,j=1\atop i<j}^N
  \beta_i\beta_j G(z_i,z_j) \cr
  & +\sum_{a=1}^s\sum _{j=1}^N \alpha \beta_j G(z_j,w_a) 
   +  \sum_{a,b=1\atop a<b}^s \alpha^2 G(w_a,w_b) + \c R \biggr \} \cr 
  }
}
Here, $Z_s(m)$ represents the functional determinant of the
scalar Laplace operator on a surface with metric $\hat g (m)$
and takes the form
\eqn\sdet{
   Z_s(m) = {{\rm Det}' \Delta _{\hat g} \over \int d^2z
     \sqrt{\hat g} } 
}
The additional term $\c R$ in \lcorr\ involves all the integrals
over the Gaussian curvature $R_{\hat g}$. Clearly, this term is
absent at string tree level where all curvature can be
concentrated at $\infty$. It is also absent at one loop level
where we can take $R_{\hat g}=0$. Later on, we shall establish
that this term is subdominant in the high energy limit and may
be omitted.

The integration over the matter fields will depend upon the
precise nature of the matter conformal field theory.
Fortunately, in anticipation of the high energy limit of the
amplitudes, we shall only need to exhibit the part of the matter
amplitudes that involves the exponential factors.
\eqn\mattercorr{
  \corru{M}{\prod_{j=1}^N V(z_j)}
    =Z_s(m)^{-d/2} \c P(z_i) \exp\biggl \{ -\sum_{i,j=1\atop i<j}^N 
    k_ik_j G(z_i,z_j)\biggr \}
  }
Here, the function $\c P (z_i)$ involves all matter Green
function factors other than exponential, and $Z_s(m)$ represents
the functional determinant of the scalar Laplace operator of
\sdet .

Combining the Liouville and matter parts of the correlation
functions, we obtain our final expression for non-critical
string correlation functions, given by the following expression 
\eqn\fullcorr{
    \corru{}{\prod_{i=1}^N \c V _i}
    =  
    {\Gamma(-s)\mu ^s \over\alpha (4\pi)^s}
      \sum_{h=0}^\infty\int_{\c M_h} \!\!\! dm\,
      Z_s(m)^{-(d-1)/2}       
    \int_{\Sigma _h}  \prod_{i=1}^Nd^2z_i \c P (z_i)
    \int _{\Sigma _h} \prod_{p=1}^sd^2w_p
    \exp\bigl \{ -\c E_h(z_i, w_p)  \bigr \}
}
The total partition function $Z(m)$ is given by
\eqn\part{Z(m)=Z_{gh} (m) Z_s ^{-(d+1)/2} (m)
}
The argument of the exponential is given by
\eqn\defi{
   \c E_h(z_i, w_p)
    =
    \sum_{i,j=1\atop i<j}^N u_{ij} G(z_i,z_j)
    - \sum_{j=1}^N\sum_{p=1}^s \alpha \beta_j G(z_j,w_p)
    - \sum_{p,q=1\atop p<q}^s \alpha ^2 G(w_p,w_q) -\c R
    }
The $u_{ij}$ are analogues of the Mandelstam variables 
extended to the
case of non-critical string theory and are given by
\eqn\sijdef{
  u_{ij}\equiv -\beta_i\beta_j+k_i \cdot k_j
  }
Momentum conservation $\sum_j k_j=0$, Eq. \eqsdef\ together with
the defining equations for $\alpha$ and $\beta$ in \kdef\ and
\vertexop, guarantee diffeomorphism invariance of the
correlation function.

It is instructive to view the function $\c E_h(z_i,w_p)$ as the
electro-static energy of an array of electric charge vectors
$K_0=(\alpha;0) $ placed at points $w_p$ and electric charge
vectors $K_i=(\beta _i; k _i)$ placed at points $z_j$. A natural
inner product may be defined on these $d+1$-dimensional vectors
in the following way
\eqn\inner{
   K_i\cdot K_j \equiv -(\beta _i + {\kappa \over 2})(\beta _j + 
   {\kappa \over
2}) + k_i \cdot  k_j
}
The on-shell condition for external states characterized by
conformal dimension $\Delta _i$ in \confvertex\ is simply reads
\eqn\square{
   K_i\cdot K_i  = -m_i ^2 
\qquad \qquad 
   m_i^2 = {1 -c \over 12} + 2 \Delta _i 
}
Thus, the computation of correlation functions in non--critical
string theory has been reduced to that of the free energy of an
array of electric charge vectors in two dimensional
electrostatics. In addition to the charges associated with
external vertex operators, which are familiar from critical
string amplitudes, the Liouville operator introduces additional,
internal, charges.

The problem of evaluating correlation functions in non-critical
string theory is seemingly reduced to the problem of computing a
finite dimensional multiple integral with respect to 
$z_i,w_a$ over the Riemann surface.  
Since $s$ is not, in general, an integer however, the
correlation function is not well defined as it stands. 
{}From the arguments presented in \AD, it is clear that the 
original expression for the amplitudes in \eqcorr\ 
is complex analytic in the external momenta and in the 
central charge $c$, even though the intermediate expressions 
\split, \lcorr\ and \fullcorr\ only make sense for integer $s$. 
Thus, the results obtained by evaluating \fullcorr\ for
integer $s$ will have to be analytically continued in $s$.
This is achieved through a combination of analytic continuation 
in the external momenta (just as in the critical string, \EDDP)
and in the central charge $c$.
 
For rational $c<1$, and to string tree level, it was proposed in
\GL\ to analytically continue in the variable $s$, using certain
rearrangement formulas for ratios of Euler $\Gamma$-functions
(that are specific to tree level). The validity of this
procedure is justified, after the fact, since it produces
agreement with results from matrix models. More importantly,
agreement can be established from first principles, as was 
shown in \AD, using a saddle point approximation in the limit 
when $\alpha\rightarrow0$, i.e. when $c\rightarrow \infty$.
We shall take these analyticity properties as a definition
for the amplitudes away from $c<1$ and rational.
\subsec{Correlation functions in the high energy limit}
For tree level amplitudes, the above multiple integrals are of
the same type as those discussed by Selberg and in \DF.
The 3-point function was obtained in their work for
arbitrary parameters, but results on the 4-point function are
limited to $c<1$ conformal matter. In general, even to string
tree level, the integrals of \fullcorr\ and \defi\ are not
available in explicit form.

We propose to evaluate the non-critical string correlation
functions in the limit where the energies and momenta of the
external string states all become large compared to the square
root of the string tension. This limit is physically interesting
and was extensively explored in the case of the critical string \GM.
Also, the limit of the amplitudes is calculable and could be
used as a starting point for a more systematic expansion in high
energy. In fact, we shall establish that the amplitudes, in the
high energy limit, are given by a saddle point, which
corresponds to the electro-static equilibrium configuration of
the associated electro-statics problem. The saddle point
configuration may be evaluated and the associated electro-static
energy --- the quantity $\c E_h$ entering \fullcorr\ --- may be
obtained explicitly, at least to tree level.

We shall define the high energy limit by rescaling all momenta
$k_i$ by a common factor $\lambda \rightarrow \infty$. We have
the following asymptotic behavior
\eqn\limit{
  k_i      \rightarrow  \lambda k_i  
\qquad \qquad 
  \beta _i \rightarrow 
  \pm \lambda |k_i |  +\c O (1) 
}
In the high energy limit, 
$K\equiv(\beta;k)$ naturally corresponds to the momentum of a
massless particle as expected.
The scaling properties of other quantities are easily deduced
from the above~: $u_{ij} $ scales like $\lambda ^2$, $s$ scales
like $\lambda ^1$ while $c$ and $\alpha$ scale like $\lambda
^0$. Notice that external vertex operators always remain
conformally invariant under this scaling.

To determine the high energy limit of the non-critical
scattering amplitude in \fullcorr, we begin by analyzing the
high energy behavior of the electro-static energy function
 $\c E_0$. {}From the expression for $\c E_0$ in \defi, it can be
 readily 
show that $\c E_h$ scales like $\lambda ^2$ for large $\lambda$.
This is manifest for the first term in $\c E_h$ in
\defi. Actually, the next two terms in \defi\ also scale like
$\lambda ^2$ for large $\lambda$. Although the couplings in the
second term only scale linearly in $\lambda$, the number of
Liouville insertion points, $s$, also grows like $\lambda$. In
the third term, the coupling $\alpha$ scales like $\lambda
^0=1$, but there are now $s^2$ Liouville insertion points, so
again this term scales like $\lambda ^2$.  The last term, $\c
R$, can be neglected in what we do. Firstly, to tree and one
loop levels, the curvature term is irrelevant. Also, its scaling
with $\lambda$ is at most linear in $\lambda$, as can be seen
from combining \split\ and \eqnonzeromodes.

The next ingredient needed in the determination of the high
energy limit of the non-critical scattering amplitudes is the
degree of dependence of this limit on any specific matter
conformal field theory. The most important simplification in
this respect comes from the observation that the conformal
primary fields $\c P _\Delta$ involve and produce only
polynomial dependence on the space-time momenta $k_i$. Thus, in
the high energy limit, where the contributions from the saddle
point will be generically exponential (as we shall establish
below), we may neglect the polynomial contributions from the
vertex functions $\c P _\Delta$. Thus, only the exponential
vertex operator parts contribute to the high energy limit.

The fact that the entire function $\c E_h$ scales in a homogeneous
way for large $\lambda$, allows us to use routine saddle point
methods to calculate the limiting behavior.\foot{In \AD, a
  saddle point approximation was applied to the Liouville
  functional integral directly to the Liouville field in the
  limit semi-classical limit where $\alpha\rightarrow0$.  In the
  high energy limit however, the Liouville action does not scale
  homogeneously and a similar procedure does not yield a good
  valid saddle point formulation.
  To summarize, the limits taken are different; in \AD, the
  external states were fixed and $c$ was taken to $-\infty$ and
  here, $c$ is fixed to be of  order one while the external
  momenta are taken  to be large.
  } 
The leading order
contribution will be given by the value of $\c E_h$ at the saddle
point, specified by the vertex positions $z_i^0$, Liouville
insertion points $w_p^0$ and moduli $m^0$ at the saddle.
\eqn\saddle{
  \corru{}{\prod _i \c V _i} = 
  \exp \{ - \c E_h \left(z_i ^0, w_p ^0; m^0\right)\}
}
The values of $z_i^0$, $w_p^0$ and $m^0$ are determined from the
saddle-point equations
\eqn\equil{
{\del \c E_h \over \del z_i}\bigg | _{z_i ^0, w_p ^0, m^0} 
= {\del \c E_h \over \del w_p}\bigg | _{z_i ^0, w_p ^0, m^0}
={\del \c E_h \over \del m} \bigg | _{z_i ^0, w_p ^0, m^0} =0 
}
Thus, the overall dependence on the momenta at high energy will
be exponential.  This saddle point equation is just the
expression for electro-static equilibrium for an array of vector
charges. In addition to the vertex operators present for the
critical string, we now also have a number $s$ of Liouville
charges.

The number of Liouville charges grows with increasing energy,
and tends to $\infty$ in the infinite energy limit. If we assume
that the saddle point Riemann surface remains compact in the
high energy limit then the Liouville charges
must accumulate somewhere on the surface. What could be the
limiting distribution of the Liouville charges ? A priori, this
distribution could be two-dimensional and fill regions of the
surface; or it could lie along line segments on the surface; or
it might be arranged in more exotic configurations, like Cantor
sets. To find out which one of these distributions is physically
realized, we shall examine the case of tree level amplitudes
first, where explicit formulas are readily obtained.

\section{Tree Level : Electro-statics on the Plane}
To tree level, the worldsheet topology is that of the sphere (or
by stereographic projection, of the complex plane), there are no
moduli, and all determinant factors $Z(m)$ are constants. The
Green function is the electro-static potential on the two
dimensional plane, given by\foot{We have made explicit in the
  definition of the Green function a short distance regulator
  $\epsilon$, which will not be exhibited in the sequel, but
  will always be subsumed.}
\eqn\gtree{
  G(z,z')= -\ln \{ |z-z'|^2 + \epsilon ^2\}
}
Tree-level non-critical amplitudes, -- evaluated for vertex
operators that are exponentials only) then reduce to a simple
multiple integral expression\foot{
Notice that these integrals resemble those evaluated in
\DF\ and \DIKUT,
but they involve more general exponents. As a result, no explicit 
formulas appear to be available for their evaluation.}
\eqn\treecorr{\eqalign{
    \corru{}{\prod_{i=1}^N \c V _i}
    &=  
    {\Gamma(-s) \mu ^s \over\alpha (4\pi )^s}
    \int  \prod_{i=1}^Nd^2z_i 
    \prod_{i,j=1\atop i<j}^N \left|z_i-z_j\right|^{2u_{ij}} 
    \cr  &\times\int \prod_{p=1}^sd^2w_p
    \prod_{j=1}^N\prod_{p=1}^s
    \left|z_j-w_p  \right|^{-2\alpha\beta_j }
    \prod_{p,q=1\atop p<q}^s\left|w_p-w_q\right|^{-2\alpha^2 }\cr
    }
  }

The associated electro-statics problem is that of an array of
charges at points $z_i$ and $w_p$, characterized by the
following electrostatic energy function
\eqn\energy{\eqalign{
\c E_0(z_i, w_p) 
=  
    - \sum _{i,j=1\atop i<j}^N u_{ij} & \ln |z_i-z_j|^2 
    +\sum _{j=1}^N\sum _{p=1}^s \alpha\beta_j \ln |z_j-w_p |^2
\cr &
    +\sum_{p,q=1\atop p<q}^s \alpha ^2 \ln |w_p-w_q|^2 \cr
    }
}
where $\alpha$ and $\beta_i$ were defined in \kdef\ and
\vertexop\ and $u_{ij} = -\beta _i \beta _j +k_i \cdot k_j$. The
saddle point equations for the integral \treecorr\ are just the
equations for electro-static equilibrium of the associated
electro-statics problem. They are given by
\eqn\treensaddle{  \eqalign{
    -\sum_{j=1\atop j\not=i}^N{2b_{ij}\over z_i-z_j}
    +{1 \over s} \sum_{p=1}^s {a_i \over z_i-w_p}  = & 0
    \cr
   {1\over s}\sum_{q\not=p\atop q=1}^s{2\over
      w_p-w_q}+\sum_{j=1}^N{\al j\over w_p-z_j} = & 0 \cr
    }
  }
Here,\foot{ The variables $\bar z _i$ and $\bar w _p$ satisfy
  \treensaddle\ with $z_i$ and $w _p$ replaced by $\bar z_i$ and
  $\bar w_p$ respectively. When the charges $a_i$ and $b_{ij}$
  are real, those respective equations are just the complex
  conjugates of one another. But when the charges $a_i$ and
  $b_{ij}$ are taken to be complex, the equations are no longer
  complex conjugates of one another, and $\bar z_i$ and $\bar
  w_p$ at the saddle point are no longer the complex conjugates
  of $z_i$ and $w_p$ respectively.  } we have defined parameters
$\al i\equiv 2\beta_i/(\alpha s)$ and $b_{ij} \equiv u_{ij} /
(\alpha s)^2$ both of which scale like $\lambda ^0$ in the limit
of large $\lambda$. Also, each summation over the number of
Liouville charges at $w_p$, $p=1,\cdots s$ has been divided by a
factor of $s$, so that the entire equations \treensaddle\ scale
like $\lambda ^0$ in the limit of large $\lambda$.

The integral \treecorr\ and the saddle point equations 
\treensaddle\
are invariant under simultaneous conformal transformations 
of $z_i$ and $w_p$. Using these transformations, we may 
fix three points
$(z_{N-2}, z_{N-1}, z_N) = (0,1,\infty)$, but we shall 
continue to denote 
0 and 1 by $z_{N-2}$ and $z_{N-1}$ respectively.

We wish to solve the equations \treensaddle\ in the limit where
$s \rightarrow \infty$, while keeping $a_i$, $b_{ij}$ and the
number of vertex charges $N$ fixed. The most difficult part of
this problem is the solution of the second equation in
\treensaddle, for the following reasons. Since the number of
Liouville charges at $w_p$ tends to $\infty$, they must 
accumulate somewhere
on the Riemann sphere. (Viewing the sphere as the complex plane,
the charges might accumulate at infinity.)  We do not know how
they accumulate and what the limiting distribution of Liouville 
charges at
$w_p$ will look like.

A priori, the limiting distribution might correspond to 
two-dimensional regions of charge, to one-dimensional 
line segments, to isolated points, or even to more 
exotic arrangements such as Cantor sets. 
We shall start by providing an answer to this question first, 
by carefully keeping the Liouville charges at $w_p$
isolated, and taking the limit only when completely safe. 
We shall find that the distribution of Liouville charges at 
$w_p$ is always of the form of a collection of curve segments, 
whose number is  $N-2$.
\subsec{Solving the Electro-statics Problem for $N=3$ by Jacobi Polynomials
}
The case of $N=3$ may be solved for any finite integer $s$ 
with the help of Jacobi polynomials, as is shown in Appendix A. 
For charges $a_1$ and $a_2$ located at $-1$ and $1$ 
(with compensating charge at $\infty$), the positions of 
the Liouville charges at $w_p$ are given by the zeros of 
the Jacobi polynomials 
\eqn\jack{
P_s ^{(-1+a_1s/2 ,~-1+a_2s/2)}(w)=0
}
For real positive $a_1$ and $a_2$, and integer $s$, it is 
clear that there are $s$ solutions $w_p$, $p=1,\cdots,s$, 
all located such that $-1 <w_p<1$. In this case, one 
easily sees that the zeros accumulate onto a line segment 
between $-1$ and $1$, as shown schematically 
in \fig\figone{}.
It is useful to keep this case in mind when generalizing 
to higher values of $N$.\nl
\vskip-2cm
\centerline{
  \epsfxsize=7.0cm
  \epsfbox{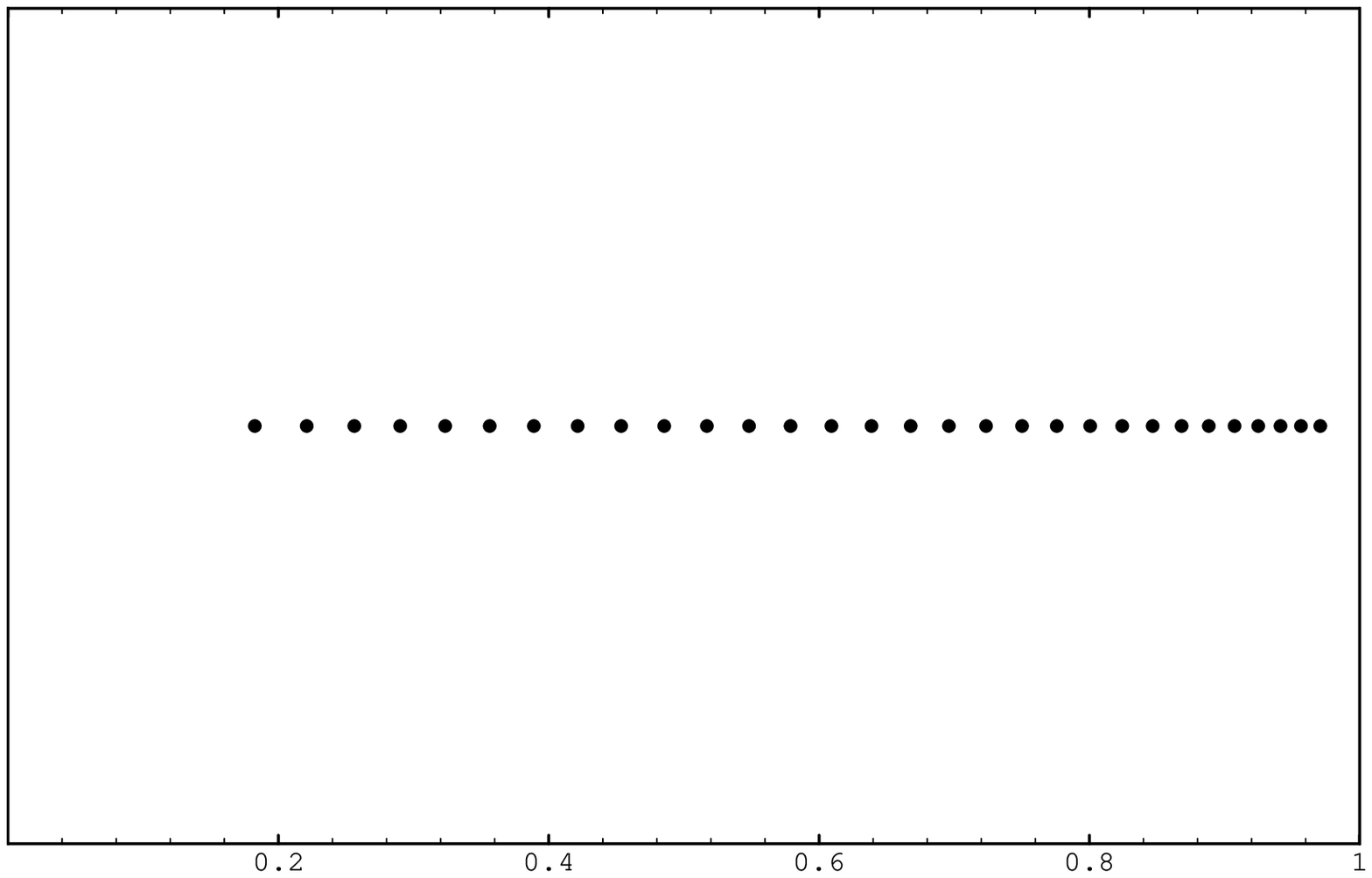}
  \epsfxsize=7.0cm
   \epsfbox{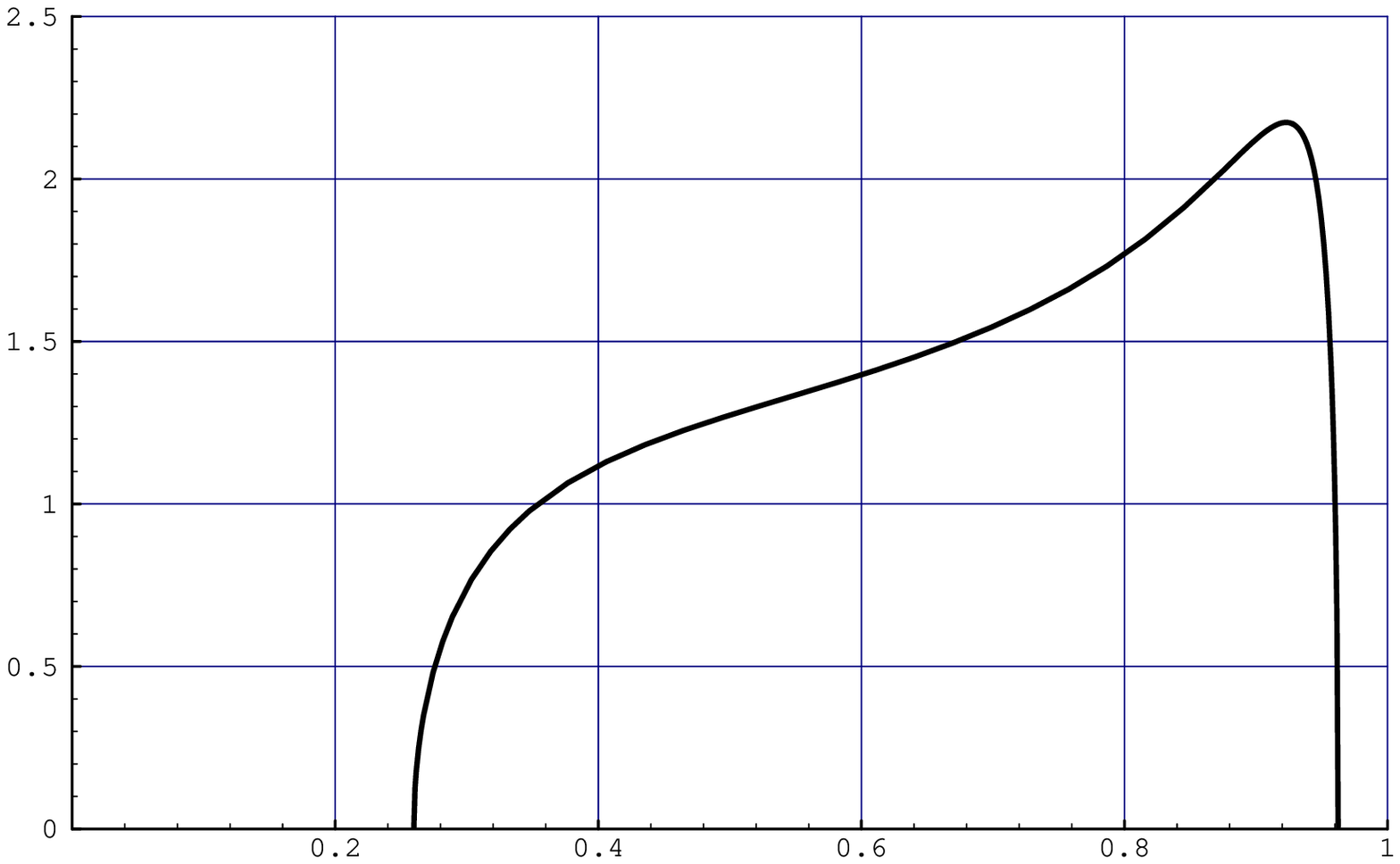}}\nl
\vskip-3.2cm
{}\hskip3cm (a) \hskip6.5cm (b)

\nl\figone(a),(b)\ {\it (a) Positions of Liouville charges $w_p\ 
  (p=1,2,\ldots,30)$  for $\al 1=3, \al2 =1$.\nl
  (b) The density of Liouville charges in the continuum limit
   for $\al1=3,\al2=1$.
}
\subsec{Solving the Associated Electro-statics Problem for 
General $N$}
To study equation \treensaddle\ for general $N$, we make 
use of a complex potential $W(z)$ for the charges $z_i$, 
defined by
\eqn\vprime{
 W(z)\equiv -\sum_{j=1}^{N-1}\al j\ln( z_j-z)
}
in terms of which the equation \treensaddle\ for the 
Liouville charges at $w_p$ becomes
\eqn\msaddle{
  {1\over s}\sum_{q=1\atop q\not =p}^s{1\over
    w_p-w_q}=\half W'(w_p) \; .
  }
This equation, for general $W(z)$ is just the electro-statics
condition for an assembly of $s$ charges in the presence of an
external potential $W(z)$ -- given here by the potential
generated by the charges at $z_i$. We also introduce a complex
analytic generating function $\omega (z)$, defined by
\eqn\res{
  \omega(z)={1\over s}\sum_{p=1} ^s {1\over w_p-z}
}
which, physically, is just the electric field produced by the 
Liouville 
charges at $w_p$. Its divergence is obtained by applying the
Cauchy-Riemann operator,  
\eqn\density{
\rho ^{(2)} (z) = -{1 \over 2\pi i} {\del \over \del \bar z} 
\omega (z) =
{1\over s} \sum _{p=1} ^s \delta ^{(2)} (z-w_p)
}
and yields the (two-dimensional) electric charge density $\rho
^{(2)} (z)$, with unit integral over the plane.

One may re-express the set of $s$ equations \msaddle\ in terms
of the following Riccati equation\foot{Similar
  methods involving generating functions were used in the study
  of matrix model spectral equations \DGZ. The potentials $W(z)$
  were typically polynomials in that case.} for $\omega(z)$
\eqn\omegaeq{
  \omega^2(z)-{1\over s}\omega'(z)+
  W'(z)\omega(z)+ {1 \over 4}R(z)=0
}
The auxiliary potential $R(z)$ is defined by
\eqn\rdef{ 
R(z) = {4\over s}\sum_{p=1}^s{W'(w_p)-W'(z)\over
    w_p-z}
  }
For general $W(z)$, it would not be possible to carry out the
sum in the definition of $R(z)$ in any simple way. When $W(z)$
is a rational function of $z$ however, as is the case here,
$R(z)$ is also rational, with poles at precisely the same
locations as $W(z)$. It is easy to determine $R(z)$
explicitly~:
\eqn\rsol{
  R(z)=\sum_{i=1}^{N-1}{R_i\over z-z_i},\qquad{\rm where}\qquad
  R_i = {4\al i \over s}\sum _{p=1}^s {1 \over w_p -z_i}
}
The fact that we have been able to determine the functional form
of $R(z)$ explicitly, in terms of a finite number of parameters
is perhaps the most important ingredient in our solution of the
associated electro-statics problem.\foot{In the case of matrix
  model spectral equations, where $W(z)$ it typically
  polynomial, the analogous key ingredient is that $R(z)$ is
  polynomial.}

All that precedes is still an exact transcription of the
electro-statics equations \treensaddle, valid for any finite
number of Liouville charges at $w_p$. We shall now bring about 
one further
simplification by using the approximation in which the number of
charges $s$ is large. (Recall that, in the original non-critical
string problem, this limit corresponds to high energy of all
external string states.)

The potential $W(z)$ is independent of $s$, while the electric
field $\omega(z)$ and the auxiliary potential $R(z)$ converge to
finite limits as $s\rightarrow \infty$. Thus, the
electro-statics equation for $\omega (z)$ of \omegaeq\ can be
simplified in this limit, as the term in $\omega '(z)$ is
suppressed by a factor of $1/s$ and may be dropped. Instead of
the Riccati equation of \omegaeq, we obtain now an algebraic
equation
\eqn\quadeq{
  \omega^2(z)+W'(z)\omega(z)+{1\over4}R(z)=0 \; .
}

This quadratic equation is easily solved, and we obtain the
electric field $\omega(z)$ of the Liouville charges at $w_p$, 
given by
\eqn\omsol{
  \omega(z)={1\over2}\left[-W'(z)\pm\sqrt{W'^2(z)-R(z)}\right],
  }
The sign in front of the square root should be chosen so that 
the poles in
$\omega (z)$, located at the points $z_i$, are absent when 
the charges $a_i$ are
all real and positive. Indeed, for charges of like sign at 
$z_i$ and $w_p$, 
no Liouville charges at $w_p$ should coincide with any of 
the $z_i$.\foot{The cases where $a_i$ are not all real and
  positive  can be obtained by
analytic continuation in the charges $a_i$. It is then in 
general possible for
Liouville charges at $w_p$ to leak into the points $z_i$, 
so that also $\omega (z)$ will then
have poles at these points.} Eq. \omsol\ also immediately 
determines the
density of charges by Eq. \density. The analysis of the 
structure of the density
will be discussed in the next subsection. 

The asymptotic behavior of the various functions yields 
simple relations between
the coefficients $R_i$, which we now determine. Since 
$W'(z)$ tends to 0 as
$1/z$ for large $z$, we see from \rdef\ that $R(z)$ 
must tend to 0 as $1/z^2$ for
large $z$. This implies that the sum of all $R_i$ must 
vanish. Furthermore,
using \omegaeq\ and the fact that $\omega (z) \sim -1/z$ 
as $z\rightarrow
\infty$, we obtain a second relation between the $R_i$'s. 
Putting all together,
we have
\eqn\aeqs{
    \sum_{i=1} ^{N-1}\  R_i =0
    \qquad\qquad
    \sum_{i=1} ^{N-1} z_i R_i = -4 -4\sum_{i=1} ^{N-1} \al i
}
The remaining $N-3$ independent parameters $R_i$ appear 
not to be determined by equation \quadeq. Their physical 
interpretation will be given in Subsection 3.3.

Since $R_N$ defined through \rsol\ satisfies $R_N\rightarrow0,
z_N R_N\rightarrow -4a_N$ when  $z_N\rightarrow\infty$, the sum in
the above  formulas may be naturally extended to be from one to
$N$, making these formulas conformally covariant.
We also note from \eqsdef\
that the sum of the charges in the high energy
limit is simply
\eqn\simplesum{
  \sum_{i=1}^N \al i=-2
}
\subsec{Structure of the Density of Liouville Charges at $w_p$}
 {}From the explicit expression in Eq. \omsol\ for the electric
field $\omega (z)$ in the limit of large $s$, we immediately
read off the distribution of Liouville charges at $w_p$ 
for the saddle
point. The function $\omega (z)$ is holomorphic 
throughout the
complex plane, except for branch cuts arising from 
the square
root in \omsol. Thus, the Liouville charges at $w_p$ 
accumulate to lie on
segments of curves that correspond to the branch cuts in the
function $\omega (z)$.

The positions of the associated branch points are 
most easily
exhibited by recasting the solution for $\omega(z)$ 
as follows~:
\eqn\polyn{
W'(z)^2 - R(z) = Q_{2N-4} (z) \times \prod _{i=1} ^{N-1} (z-z_i)^{-2}
} 
Here, $Q_{2N-4}(z)$ is a polynomial in $z$, which is of degree
$2N-4$, in view of the fact that the sum of all $R_i$ vanishes,
as shown in \aeqs. Using also the second equation in \aeqs,
we find
\eqn\polnorm{
Q_{2N-4} (z) = (a+2) ^2 \prod _{k=1} ^{2N-4} (z-x_k) 
\qquad\qquad
a = \sum _{i=1} ^{N-1} a_i
}
The function $\omega (z)$ thus exhibits $N-2$ branch cuts, 
$\c C_p$, spanned between pairs of branch points $x_{2p-1}$ 
and $x_{2p}$, $p=1,\cdots ,N-2$, of $\omega(z)$, which 
correspond to zeros of the polynomial $Q_{2N-4}(z)$. 
Next, we shall address the more detailed issue as to 
exactly where the charges lie.

Since the configuration of the Liouville charges at 
$w_p$ is one-dimensional,
it is convenient to use a notation where this fact 
is clearly
brought out. Since the support of the Liouville 
charge density is
a collection of curve segments, $\c C = \c C _1 
\cup \cdots \cup \c C_{N-2}$, the two-dimensional 
density of charges $\rho ^{(2)}(z)$ of \density\ 
can be rewritten in terms of a line density of 
charges $\rho
(z)$ as follows
\eqn\rhodef{\eqalign{
\rho ^{(2)} (z) = & \int _{\c C} dw ~ \rho (w) \cr
                = & \int dt ~\dot w(t) ~\rho (w(t)) 
                ~\delta ^{(2)} (z-w(t))\cr
 }
}
Here, the curve segments of $\c C$ are parametrized 
by a real parameter $t$, and $\dot w$ denotes the 
$t$-derivative of $w$. Given the solution for $\omega (z)$
in \omsol, it is straightforward to calculate the 
linear density
$\rho(w)$, defined when $w$ lies on $\c C$.
\eqn\rhodensity{
\rho (w) = { 1 \over 2\pi} \sqrt {R(w)-W'(w)^2 } 
}
The requirement that $\c C$ lie along branch
cuts of $\omega$ does not determine the precise position of $\c
C$. In fact, any analytic curve that joins pairs of branch
points would do. To find the saddle point distribution of the
charges, an additional ingredient must be clarified.

{}From the fact that the Liouville charges at the 
points $w_p$ are all of
unit strength times $1/s$, it follows that the 
charge density must be
real and positive along $\c C$. This supplementary 
condition requires that
the position of the branch cut $\c C$, supporting 
the Liouville charges at $w_p$, must be such that 
$dw \rho(w)$ is real as $w$ is varied along $\c C$. 
Parametrizing $\c C$ again by a real parameter $t$, 
we have 
\eqn\reality{
\overline{ \dot w (t) \rho (w(t))   }  =  \dot w (t) \rho (w(t))  
}
Equivalently, this condition may be expressed in terms 
of the (Abelian) integral\foot{
A more complete study of these integrals will be 
carried out in Subsection 3.5.}
 associated with the (Abelian) differential $dw \rho(w)$ :
\eqn\abelreal{
        \overline{I(w(t))}=  I(w(t)) 
                \qquad \qquad
        I(w) = \int _{x_1} ^w dv \rho (v)
}
When $\alpha$ and $\beta_i$ are real and positive , this simply
implies that the Liouville charges at $w_p$ are concentrated on
the real axis, as was expected. However, \reality\ and
\abelreal\ also provide consistent prescriptions for the case
when $\alpha$, $\beta _i$ as well as the external momenta, are
analytically continued to complex values.

We conclude this subsection by providing a 
semi-quantitative description of the locations of the 
branch points and branch cuts of $\omega (z)$. 
The simplest case is when all the parameters $a_i$ 
are real and positive, and when all $z_i$ are real 
as well. On physical grounds, and by symmetry arguments, 
the Liouville charge density lies on the real axis, 
and consists of $N-2$ line segments $\c C _p$, 
$p=1,\cdots ,N-2$, located in between the charges 
at $z_i$. The charge segments do not touch the points 
$z_i$. This configuration is schematically represented 
in \fig\figtwo{}.\nl\nl
\centerline{\epsfysize=6cm
  \epsfbox{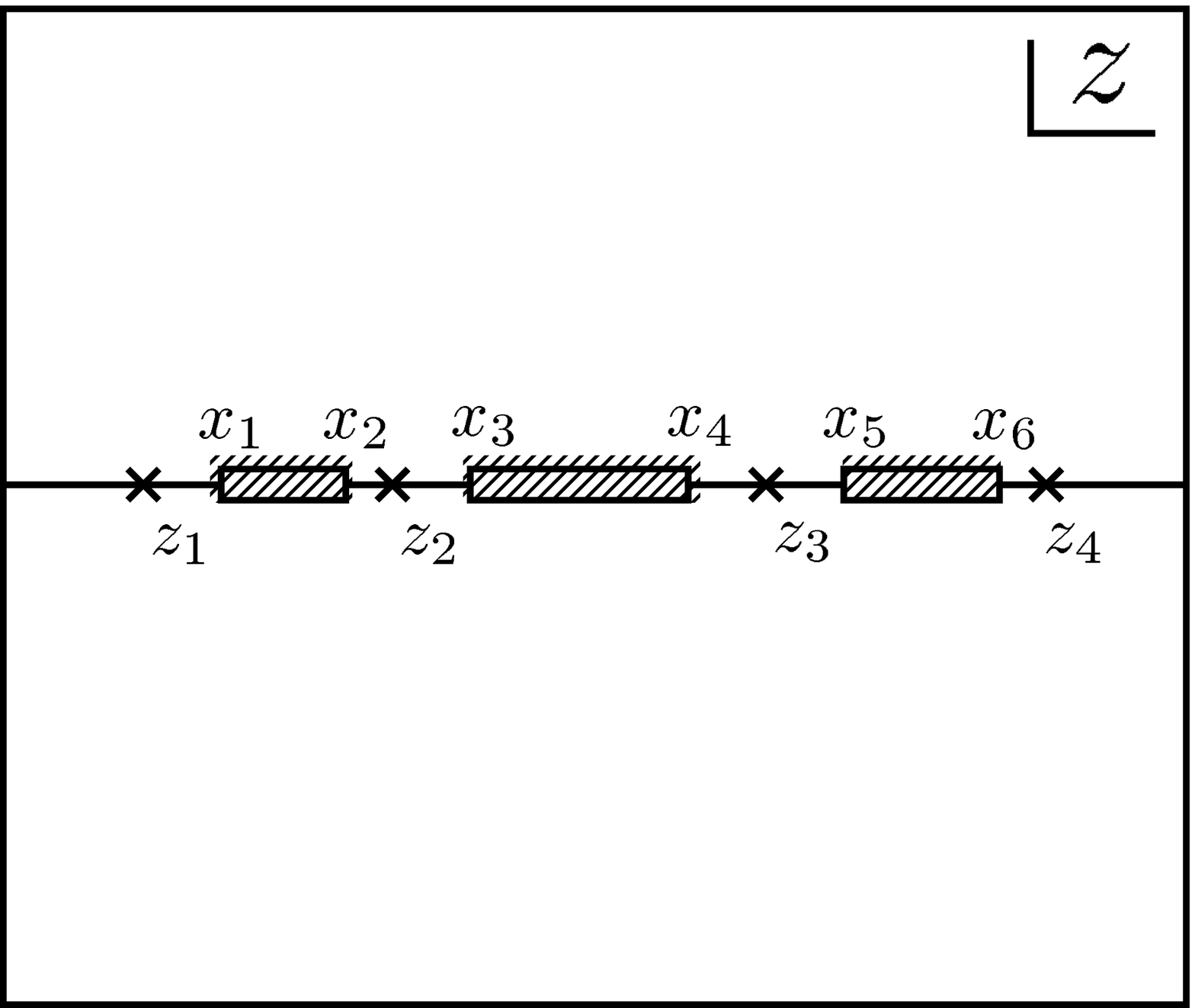}}\smallskip\noindent
\nl\figtwo\ {\it Schematic representation of the positions of the
branch points and the branch cuts for $N=5$ and real charges $\al i
(i=1,2,\ldots,5)$. }\nl\nl
On the other hand, when the parameters $a_i$ and 
the points $z_i$ are  complex, the branch points 
move out into the complex plane, away from the 
real axis. The positions of the branch points 
are given by roots of polynomials of degree 
$2N-4$, and cannot, in general, be exhibited 
explicitly. Qualitatively however, and using 
analyticity arguments, we may continuously 
turn on the complex parts of the parameters 
$a_i$ and $z_i$, and follow the branch points 
and branch cuts as they move off into the complex 
plane. This effect is represented schematically 
in \fig\figthree{}.\nl\nl
\centerline{\epsfysize=6.0cm\epsfbox{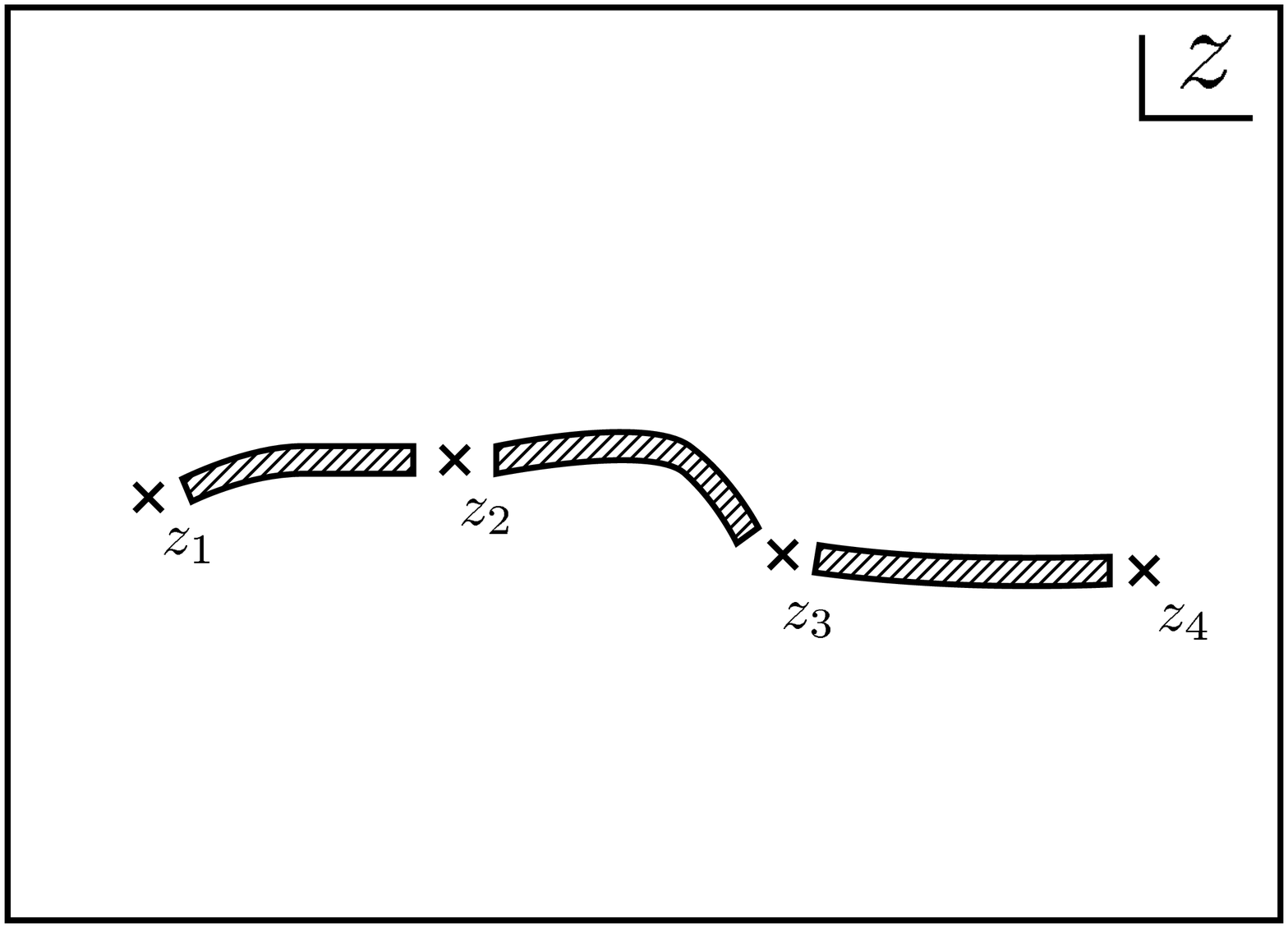}}\smallskip\noindent
\nl\figthree\ {\it Schematic representation of the positions of
  the branch points and the branch cuts for $N=5$, but $\al i $
  and complex charges $z_i (i=1,2,\ldots,5)$.
}\nl\nl
\subsec{Closure of Equations}
It is convenient to re-express the basic equations we have
obtained for the charges at $w_p$ and $z_i$ in terms of the
line density $\rho (w)$. With the help of equation \vprime\ 
defining $W(z)$, we have the following expressions for $\rho(w)$
\eqn\summary{\eqalign{
    \rho (w)
    & =
    {1 \over 2\pi} \sqrt {-Q_{2N-4} (w)} \prod _{i=1}^{N-1} 
    {1 \over w-z_i}
    \cr
    Q_{2N-4}(w) 
    & =
    \left \{ W'(w)^2 - \sum _{i=1} ^{N-1} {R_i \over w-z_i} 
    \right
    \} \prod _{i=1} ^{N-1}     (w-z_i)^2
    \cr
}
}
The total Liouville charge, obtained by integrating 
along all the curve segments of $\c C$, is normalized to 1~:
\eqn\rhonorm{    
    \int _{\c C} dw \rho (w)  =1
  }
The electric field $\omega (z)$ is defined by
\eqn\omdef{\eqalign{
\omega (z) = & \int _{\c C} dw ~\rho (w) {1 \over w-z} \cr
           = & -\half W'(z) \pm \half \sqrt {Q_{2N-4}(z)} 
           \prod _{i=1} ^{N-1}
           {1 \over z-z_i}\cr}
}
It satisfies the following equations, which are obtained as the
$s\rightarrow \infty$ limit of equations \msaddle\ and
\density. 
\eqn\contsaddle{\eqalign{
    W'(w) & = -\{ \omega (w^+) + \omega (w^-) \}  = 2 \pint
    _{\c C} dw' \rho (w') 
    {1 \over w-w'} \cr
    \rho (w) & = {1 \over 2\pi i} \{ \omega (w^+) - \omega (w^-) \}
    \cr 
    }
  }
Here, the integral is performed with principal value
prescription, so that $w^\pm$ are taken just above 
and just
below the curve $\c C$ of Liouville charges; the 
above equation for
$\rho(w)$ holds for $w$ on $\c C$.

It remains to clarify the physical significance of 
the constants $R_i$, which enter the function 
$R(z)$ in \rsol\ and the polynomial $Q_{2N-4}(z)$ 
in \polyn. By construction, they are given via the 
$s \to \infty$ limit of  \rsol, which can be 
expressed in terms of $\omega (z_i)$~:
\eqn\summaryom{ 
    R_i 
      = 
    4a_i \int _{\c C} dw \rho (w) {1 \over w-z_i}
    = 4a_i \omega (z_i) 
  }
These relations are {\it automatically} satisfied 
by the construction of $\omega (z)$, as can be 
checked easily by taking the limit of \omdef\ 
when $z\to z_i$. It thus appears that the $N-3$ 
independent parameters $R_i$, entering the 
solution of the second set of equations in 
\treensaddle, for the position of the Liouville 
charges at $w_p$, are undetermined by these equations. 

How can this indeterminacy be understood ? 
It is easiest to analyze first the case where 
all $z_i$, and all $a_i$ are real. By construction, 
the $R_i$ are then real, as can be seen from 
\rsol\ and \summary. When all 
$a_i$, $i=1,\cdots, N-1$ are positive, (and only 
the compensating charge at $\infty$ is negative) 
the possible locations for the Liouville charges 
are on the $N-2$ line segments $\c C_p$, 
$p=1,\cdots , N-2$, in between pairs of consecutive 
positive charges at $z_i$. However, exactly how the 
total Liouville charge (which is fixed to be 1) is 
partitioned among the $N-2$ line segments is not 
\`a priori determined. Indeed, the positive 
Liouville charges cannot cross over from one 
line segment into another,
since crossing would involve passing through a 
charge configuration of infinite electro-static 
energy when a Liouville charge is on top of a 
charge $z_i$. Thus, for any partition of the 
Liouville charges among the $N-2$ intervals, 
there must be an equilibrium configuration, 
and the $N-3$ independent parameters $R_i$ 
precisely specify the possible partitions of 
the Liouville charges over the $N-2$ intervals. 

When the points $z_i$ move into the complex plane, 
and the charges $a_i$ are allowed to become complex 
as well, the allowed line segments on which the 
Liouville charges can lie move into the complex 
plane and become more general curve segments. 
Reality of the Liouville charges continues to 
impose the constraint that the Abelian integral
of \abelreal\ is real. In particular, the complete Abelian integrals
(or A-periods) encircling any given branch cut, must always be
real. This condition will be made more explicit in sect. 3.6 : it
imposes $N-3$ reality relations between the complex variables $R_i$.
The remaining $N-3$ free parameters in $R_i$ specify the partition of
the Liouville charges among the various curve segments, just as was
the case for real $a_i$ and $z_i$. To summarize, $N-3$ real relations
exists amongst the $N-3$ complex parameters $R_i$, and the remaining
$N-3$ real parameters of $R_i$ are undetermined and specify the
partition of the Liouville charges.

In fact, the values of the parameters $R_i$ are 
determined by the first set of equations in 
\treensaddle, which give the positions of the 
external vertex charges $z_i$, $i=1,\cdots, N-1$. 
Expressing them in terms of the high energy limit, 
where $s\to \infty$, we have
\eqn\firsteq{
\sum_{j=1\atop j\not=i}^{N-1}{2b_{ij}\over z_i-z_j} 
+{1 \over 4} R_i =0
}
Notice that the conditions in \aeqs\ are
automatically satisfied in \firsteq, in view of the symmetry of
$b_{ij}$, momentum conservation and \eqsdef. 
Therefore, out of the original $N-1$ equations in \firsteq, two
correspond to the asymptotic conditions \aeqs, leaving $N-3$
equations. In view of the analysis of the previous paragraph, only
$N-3$ real parameters amongst the $N-3$ complex $R_i$ are determined
by the electrostatics equations and the reality conditions, leaving
$N-3$ real parameters undetermined. The saddle point equations for the
correlation function of non--critical string theory in the high
energy limit comprise of
\contsaddle\ and
\firsteq.
\subsec{Electro-static Energy}
The tree level correlation function, in the saddle point
approximation, is given by 
\eqn\treefinal{
    \corru{}{\prod_{i=1}^N \c V _i}    =  
    {\Gamma(-s) \mu ^s \over\alpha (4 \pi )^s}
      e^{-{\c E}_0}
      }
where all the quantities are to be evaluated at the 
saddle point. In particular, the electro-static 
equilibrium energy is given in terms of $\rho (w)$ 
and $W(w)$ by
\eqn\sdef{\eqalign{
    -{\c E _0\over2(\alpha s)^2}
   = 
    \half \sum_{{i,j=1\atop i\not=j}}^{N-1}   b_{ij} &
      \ln  |z_i-z_j |^2
     + \int _{\c C} dw \rho(w) \{ W(w)  + \bar W(w) \} \cr
   & - \half \int _{\c C} dv\rho (v)\int _{\c C} 
   dw\rho (w)\ln |v-w|^2
 \cr }
}
Now, there is a very important simplification that can be
administered to this expression. The key observations were already 
made previously. First, the saddle point equations for $\bar z_i$
and $\bar w_p$ are the same as for the quantities $ z_i$ and
$ w_p$, even when the charges $a_i$ and $b_{ij}$ are
complex. Second, the integration measure $dw\rho (w)$ must
be real along the line segments of charge density, as pointed
out in \reality.

We see that, as a result, the entire electro-static energy is a
sum of a contribution from $z_i$ and $w_p$ on the one hand, and
the same functional form, evaluated on $\bar z_i$ and $\bar w_p$
on the other hand. Thus, given the identity of the equations for
barred and unbarred quantities, the electro-static energy is
just twice that evaluated on unbarred quantities only~:
\eqn\sdefcom{\eqalign{
    -{\c E _0\over2(\alpha s)^2}
   = 
     \sum_{{i,j=1\atop i\not=j}}^{N-1}   b_{ij} &
      \ln  (z_i-z_j ) 
     + 2\int _{\c C} dw \rho(w)  W(w)  \cr
   & -  \int _{\c C} dv\rho (v)\int _{\c C} dw\rho (w)\ln (v-w)
 \cr }
}
This simplified expression for the electro-static energy at 
equilibrium has the advantage that it has been recast in 
terms of complex analytic integrals only, involving $z_i$, 
$\rho(v)$ and $\ln (v-w)$.

To make this reformulation more explicit, we introduce the
holomorphic potential  
\eqn\phidef{
  \Omega (z) = \int _{\c C} dw\rho(w)\ln(z-w)
  }
of the Liouville charges at $w_p$. This potential is the 
analogue of the
potential $W$ for the charges $z_i$, and its derivative is
$\omega (z) = \Omega (z)'$. In terms of this function, we may
evaluate the electro-static potential of the $w_p$-charges in a
simplified way. We begin with\foot{
We arrange the logarithm so that $\overline{\ln (z-w) } 
= \ln (\bar z - \bar w)$.}
\eqn\poten{
\Re \Omega (z) = \half  \int _{\c C} dw\rho(w)\ln |z-w|^2 
}
By the electro-static equilibrium equation \contsaddle\ 
for the Liouville charges at $w_p$, its derivative is 
related to $W'(v)$, when $v$ is on the curve ${\c C}$ :
\eqn\potender{
2{\del \over \del v} \Re ~ \Omega (v) =  \pint _{\c C} 
dw\rho(w){1 \over v-w} = \half W'(v)}
Upon integration along the curve ${\cal C}$, we find that 
\eqn\integr{
\Omega (v) =  \half W(v)  + W_0 
}
where $v$ is on $\c C$, and $W_0$ is a complex constant. 
An interesting relation for
$W_0$ is obtained by summing \integr\ over the zeros $x_k$,
$k=1,\cdots , 2N-4$ of the polynomial $Q_{2N-4}(v)$. We find
that 
\eqn\w{\eqalign{
(2N-4) W_0  = \sum _{i=1} ^{N-1} & a_i \ln a_i -(a+2) \ln (a+2)
+ \sum _{i=1} ^{N-1} a_i \sum _{j\not=i} \ln (z_i - z_j) \cr
& +  \int _{\c C} dw~ \rho (w) \ln Q_{2N-4} (w)  \cr
}
}
As a result, the electro-static energy may be re-written in
terms of a single integration over $w$
\eqn\simplify{
     - {\c E _0\over2(\alpha s)^2}
   = 
    \sum_{{i,j=1\atop i\not=j}}^{N-1}   b_{ij} 
      \ln  (z_i-z_j ) -    W_0 
    -   \sum _{i=1}^{N-1} a_i \Omega (z_i)
}
These quantities are now all holomorphic, and as such will not
be changed upon continuous changes in the curve $\c C$. Thus,
any curve $\c C$, connecting the branch points can be used in
the expression above, which greatly simplifies its
calculability.
\subsec{Solution in Terms of Hyper-Elliptic Integrals}
The above electro-statics problem of charges $z_i$ and $w_p$ 
on the plane, with $N-2$ quadratic branch cuts, is naturally 
reformulated in terms of Abelian integrals on an associated 
hyperelliptic Riemann surface $\Sigma$ of genus $N-3$. 
The surface is most easily defined by an algebraic 
equation in ${\bf C} \times {\bf C}$ (or more accurately 
in ${\bf CP}^2$), given by
\eqn\curve{
y^2 = Q_{2N-4} (w) = (a+2)^2\prod _{k=1} ^ {2N-4} (w-x_k)
}
The branch cuts, associated with the curve segments $\c C _p$, 
(spanned between pairs of branch points $x_{2p-1}$ and $x_{2p}$) 
$p=1,\cdots , N-2$, may be double covered, and used as a 
basis for the ``A-cycles'' of the surface $\Sigma$, denoted by 
$A_p$. (The cycle $A_{N-2}$ also enters in our calculations 
since we deal with punctures on the surface.) The remaining 
curve segments, (spanned between the pairs of branch points 
$x_{2p}$ and $x_{2p+1}$) $p=1,\cdots , N-2$, may also be 
doubled on the second sheet of $\Sigma$ and used as a basis 
for the ``B-cycles'' of the surface $\Sigma$, denoted by $B_p$, 
$p=1, \cdots , N-3$. This construction is schematically 
represented in \fig\figfour{}.\nl\nl
\centerline{
  \epsfysize=10.0cm\epsfbox{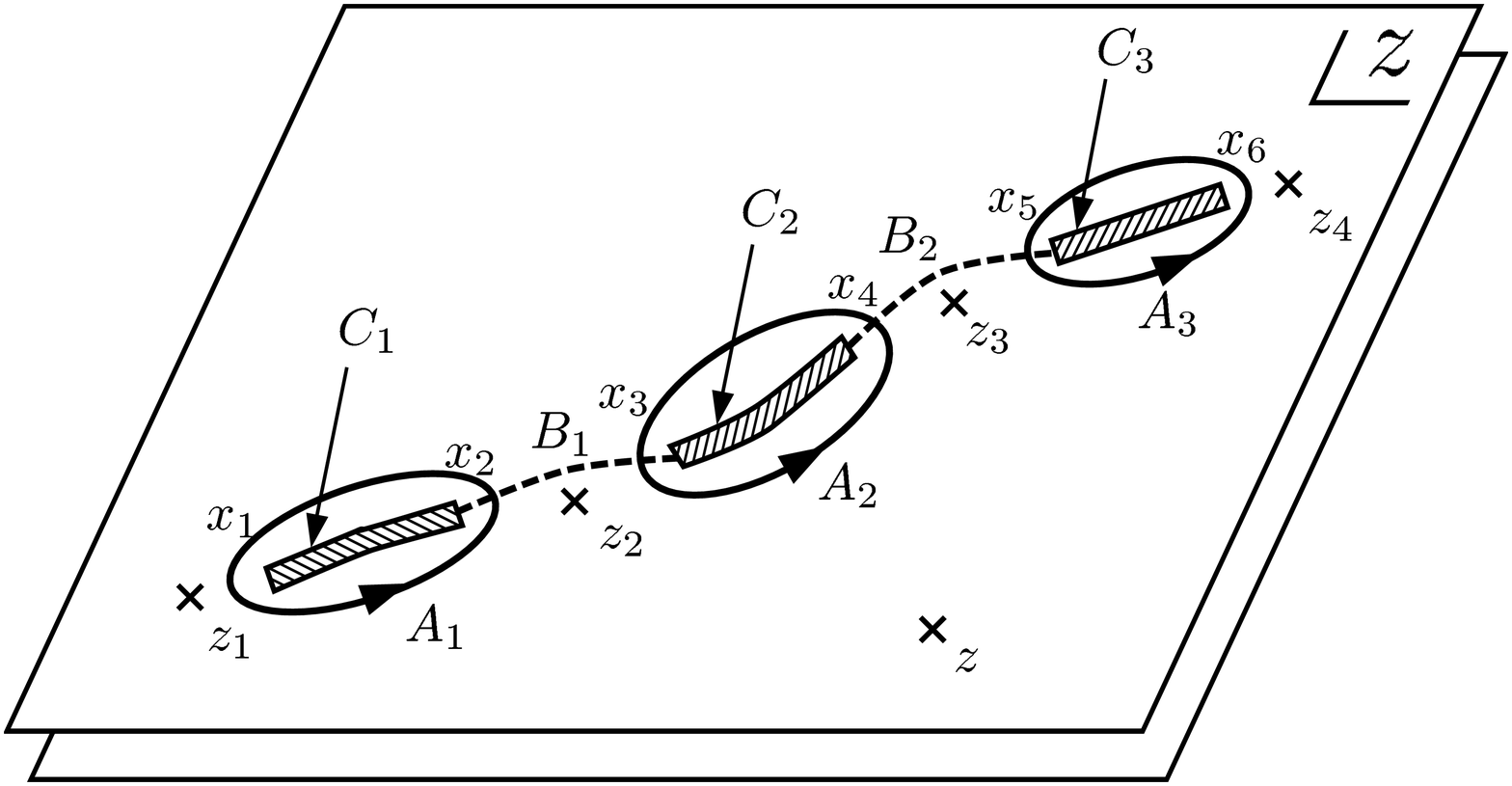}}\nl
\vskip-4.5cm
\centerline{(a)}\nl
\vskip1cm
\centerline{
  \epsfysize=8.0cm\epsfbox{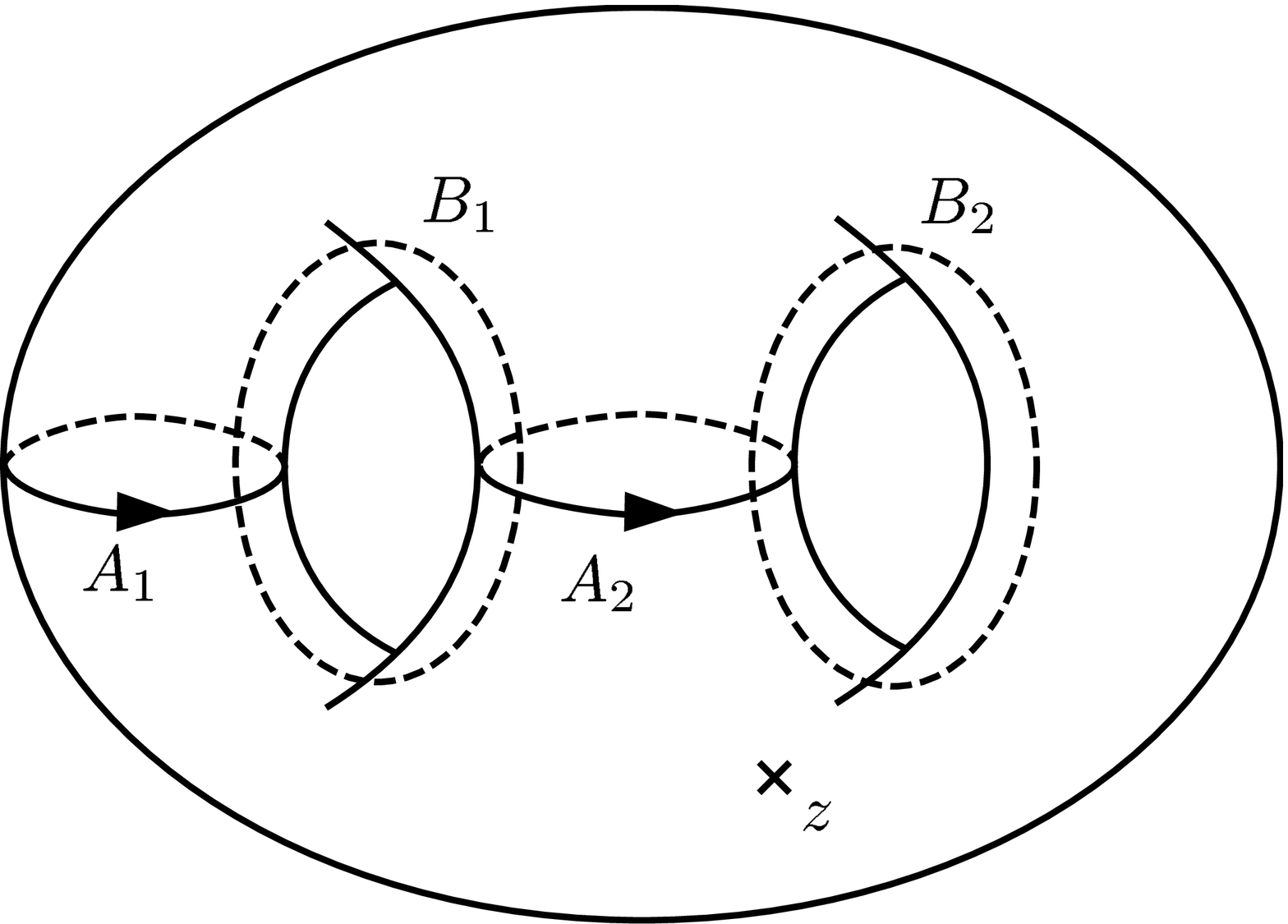}
  }\smallskip\noindent
\centerline{(b)}
\nl\figfour\ {\it Double cover of the complex $w$--plane with
  branch cuts (a) and its associated hyperelliptic Riemann
  surface (b).
}\nl\nl

On the cut sphere, the one form $\rho (w) dw$ has simple poles
at the points $z_i$, $i=1,\cdots , N-1$ with residues $a_i/(2\pi
i)$ and a simple pole at $\infty$ with residue $-(a+2)/(2\pi
i)$; away from these $N$ poles, $\rho(w)dw$ is holomorphic.  The
Liouville line charge density $\rho (w)$ naturally defines a
meromorphic Abelian differential 1-form, $\rho (w)dw$, on the
Riemann surface $\Sigma$, with the following properties. $\rho
(w) dw$ is holomorphic on $\Sigma$ apart from the simple poles
at the $2(N-1)$ points corresponding to $z_i$, $i=1,\cdots ,
N-1$ with the residues $\pm a_i/(2\pi i)$ and at the two points
corresponding to $\infty$ with the residues $\mp(a+2)/(2\pi i)$.
Thus, the differential $\rho(w)dw$ must be a superposition of
Abelian differentials on $\Sigma$ of the first and the third
kind.
More concretely, denoting the $2N$ points that correspond to
$z_i \ (i=1,2,\ldots,N)$ on the Riemann surface itself as
$z_i^\pm$, 
\eqn\rhois{ dw\rho(w)=\sum_{i=1}^N \al i \tau_{z_i^+z_i^-}
  +\sum_{p=1}^{N-3}\gamma_p\omega_p }
where $\omega_p$'s are the abelian differentials of the first
kind normalized in the canonical fashion
\eqn\abdiffnorm{
  \int_{A_p}\!\!\omega_q=\delta_{pq},\qquad
  \int_{B_p}\!\!\omega_q=\Omega_{pq},\qquad
  p,q=1,2,\ldots,N-3
}
$\left(\Omega_{pq}\right)_{p,q=1,2,\ldots N-3}$ is called the
period matrix of the 
Riemann surface.
$\tau_{xy}(z)$ is the Abelian differential of the
third kind with simple poles at $x,y$ with residues
$1,-1$
\eqn\primeform{
  \tau_{xy}(z)\equiv dz{\partial\over\partial z} \ln{
    \c E(z,x)\over \c E(z,y)}
  = dz{\partial\over\partial z} \ln{
    \vartheta[b](z-x)\over\vartheta[b](z-y)}
  }
Here, $\c E(x,y)$ is the prime form on the Riemann surface and
$b$ is any odd half period.
$\tau_{xy}$ satisfies the normalization condition
\eqn\threenorm{
  \int_{A_p}\!\!\tau_{xy}=0,\qquad p=1,2,\ldots,N-3
}
The reality condition \reality\ corresponds to  
\eqn\gammareal{\gamma_p\in\IR,\qquad p=1,2,\ldots,N-3}
The $\gamma_p$'s may be completely determined by computing the
charge in each line segment, namely
\eqn\gammafinal{
  \gamma_p=2\int_{x_{2p-1}}^{x_{2p}}\!\!\!dw\,\rho (w)
  }
Since only $N-3$ of the $A_p$ cycles are homologically
independent, the condition on the integral of $\rho(w)$ on
$A_{N-2}$ is determined as a consistency condition from the
total charge \simplesum.

The physical requirement that the density $\rho(w)$ 
should represent real, positive Liouville charges, 
demands that $\rho (w)dw$ be real along the equilibrium 
configuration of the curve segments $\c C_p$. As a result, 
the contour integrals of $\rho (w)dw$ along any $\c C_p$, 
or equivalently, along any $A_p$-cycle, must be real. 
In addition, with the proper orientation (clockwise), 
the integrals should be positive and add up to 1 along 
the curves $\c C_p$, or 2 along the closed cycles $A_p$.
\eqn\rhopos{
\int _{A_p} \rho (w) dw \geq 0
\qquad \qquad
\sum _{p=1} ^{N-2} \int _{A_p} \rho (w) dw =2
}
Of course, this requirement of reality will follow 
as soon as the reality conditions \reality\ and \abelreal\ hold.
However,  the two reality conditions have a somewhat different 
interpretation. The integral in \rhopos\ is unchanged 
under continuous deformations of the cycles $A_p$, 
and thus gives a restriction on the differential 
$\rho (w) dw$ itself, specifically on the parameters 
$R_i$, as explained in sect. 3.4. Thus, eq. \rhopos\ is a necessary
condition  on the differential $\rho (w)dw$, which must be 
satisfied if \reality\ and \abelreal\ are to hold for any curve
segments $C_p$.  Then, once \rhopos\ is satisfied, the proper 
interpretation of \reality\ and \abelreal\ is that 
they determine the curves $\c C _p$, 
within the general homology class $\half  A_p$, 
such that also the local charge density corresponds to real 
positive Liouville charges.

Next, we study the integrals in \omdef\ and \phidef\ 
defining $\omega (z)$ and $\Omega (z)$, since they 
are required in the calculation of the electro-static 
energy. It may, at first, appear surprising that 
the integral \omdef, for the electric field $\omega (z)$, 
yields an algebraic  result, even though it involves 
Abelian integrals over the Abelian differential $\rho (w)dw$. 
The reason that the result is algebraic can be understood 
from the fact that the integration contour  $\c C$ 
surrounds all branch cuts at once. Thus, by contour 
deformation, $\c C$ may be unfolded onto a sum of the 
pole contributions at points $z_i$, which may be 
carried out in an elementary way, yielding algebraic 
results (see Fig. 4 for the configuration of branch 
cuts and poles).\foot{
Note that the contributions from the contour at 
$\infty$ vanish here.}

By a similar type of contour deformation, we may 
also simplify the calculation of the integral  
\phidef, defining $\Omega (z)$, and reexpress it 
in terms of Abelian integrals only. Since the 
contributions on the contour from $w \to \infty$ 
do not vanish in this case, we shall have to proceed 
with additional care here. To do so, we introduce 
an auxiliary point $z_0$, which we may take to 
$\infty$ in the end. (see \fig\figfive{}.)
\nl\nl
\centerline{
  \epsfysize=11.0cm
  \epsfbox{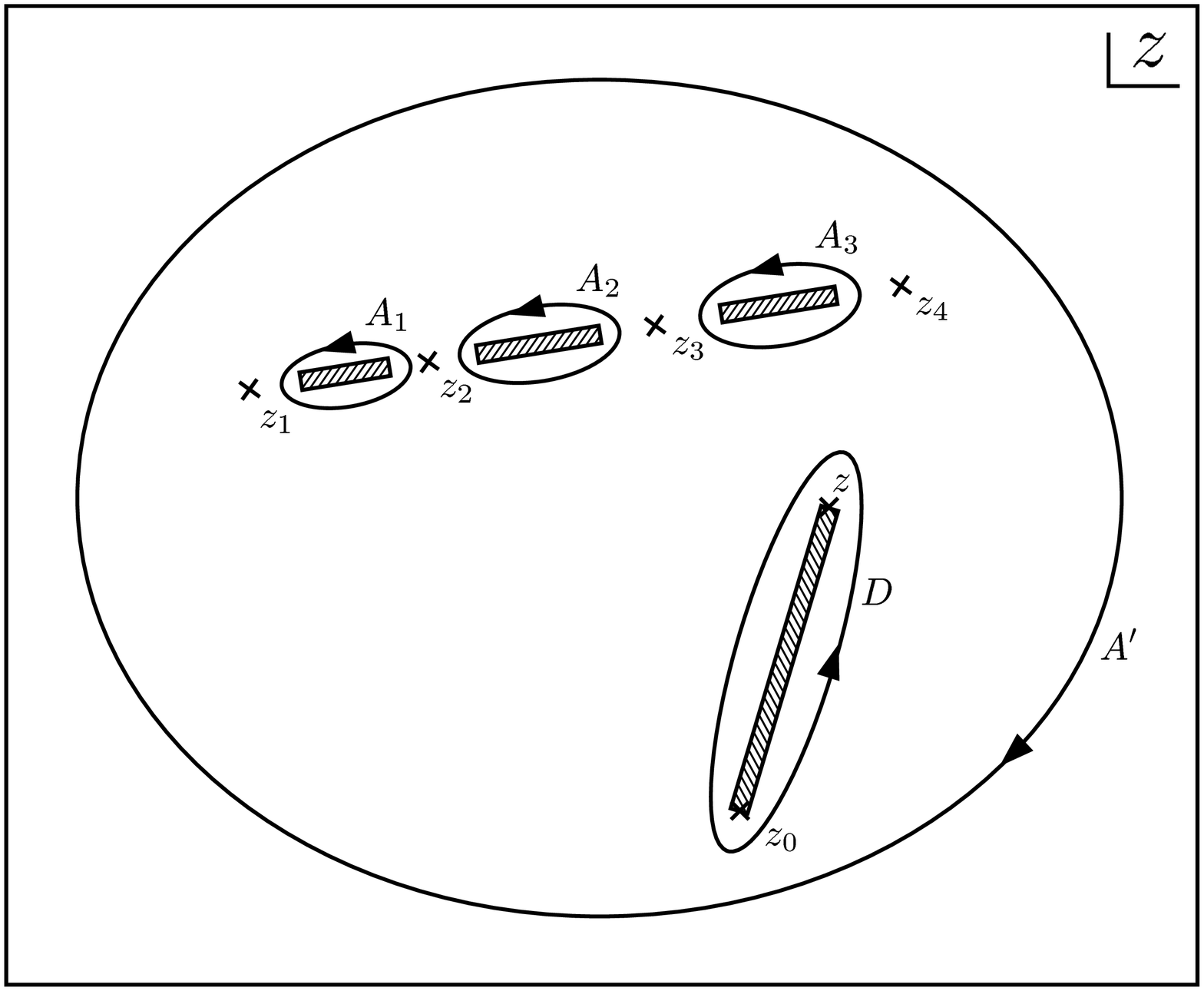}}
\nl\figfive\ {\it Complex $w$--plane with branch cuts due to
  Liouville charges and due to logarithm for the evaluation of
  $\Omega(z)$. 
}\nl\nl
 We evaluate  the integral around the sum of all 
 $A_p$-cycles : $A=A_1 + \cdots + A_{N-2}$, and 
 consider the difference integral
\eqn\difomega{
\Omega (z) - \Omega (z_0) = \half \int _A dw 
\rho (w) \ln { z-w \over z_0 -w}
}
This integration may be unfolded onto contributions 
from the poles at $z_i$ and the contour $D$ around 
the logarithmic cut (see Fig. 5). The contribution 
from the contour $A'$ vanishes, since it converges 
to zero in the $\infty$ radius limit. Thus, we obtain 
a formula reexpressing the difference of $\Omega$ at 
$z$ and $z_0$ in terms of an Abelian integral :
\eqn\abelom{
\Omega (z) - \Omega (z_0) = - \half \sum _{i=1} ^{N-1}
 a_i \ln { z-z_i \over z_0 -z_i} - i \pi \int _{z_0} ^z 
 dw \rho (w)
}
We recover the expression for $\Omega (z)$ alone by 
considering the limit $z_0\to \infty$, where 
$\Omega (z_0) \sim \ln z_0 + \c O (1/z_0)$
and we have
\eqn\omresult{
\Omega (z) = \half W(z) - i \pi \int _{z_0} ^z dw 
\rho (w) + \half (a+2) \ln z_0
\qquad \quad {\rm as} ~ z_0 \to \infty
}
This formula reproduces the result, derived in 
\integr, that $\Omega$
and $\half W$ are equal up to a constant, as $z$ 
is restricted to lie on $\c C$.
To see this, notice that the contribution of the 
Abelian integral of $\rho$ is purely real as $z$ 
is varied along $\c C$, in view of \abelreal.
It is easy to see that such integrals are trigonometric when
$N=3$ (the three-point function), elliptic when $N=4$ (the
four-point function), and hyper-elliptic when $N\geq 4$. 
The tree level correlation function of non-critical string
theory in the high energy limit is obtained as \treefinal, where 
$\c E_0$ is 
\eqn\ezerofinal{
  \eqalign{
    -{\c E_0\over2(\alpha s)^2} &= \sum_{i,j=1\atop j\not=i}^{N-1}
    b_{ij}\ln(z_i-z_j)-\sum_{i=1}^{N-1}\al i\Omega(z_i)
    \cr&\quad
    -{1\over2N-4}\left[\sum_{i=1}^{N-1}\al i\ln\al i
      -a\ln(a+2)
      +\sum_{i=1}^{N-1}\al i\sum_{j=1\atop j\not=i}^{N-1}
      \ln(z_i-z_j) + \sum_{l=1}^{2N-4}\Omega(w_l)\right]
    \cr}
  }
In view of \omresult, it remains to perform only
the integral of the density function.
In Sect. 4, we shall produce explicit formulas for the cases
$N=3,~4$.
\section{Tree Level Three- and Four-Point Functions}
To illustrate the general methods developed in the previous
sections, we shall here compute the tree level 3- and 4-point
amplitudes, in the high energy limit. Actually, the three point
function is already known~: it was calculated \GL\ for rational
conformal field theories with $c<1$, and it was shown \AD, by
analytic continuation, that this formula extends to all values of
$c$. Thus, the calculation of the three point function will
provide a non-trivial check that our methods are indeed
consistent with known result.

The 4-point amplitude, on the other hand, is not known for
general values of $c$, because of the following reasons. For
rational conformal field theories with $c<1$, the 4-point
function is known~: it satisfies a generalized hypergeometric
differential equation. The order of this differential equation
varies with $c$ in a discontinuous way~: if $c= 1 -6
(p-q)^2/(p+q)^2$, the order is essentially $p+q+1$, which varies
discontinuously with $c$. In particular, unless $c$ is rational,
the order of the equation is in fact infinite. This is why the
expressions for the 4-point function for rational conformal
field theories cannot, in any direct way, be continued away from
$c$ rational. We shall show below, however, that in the high
energy limit, our formalism allows us to evaluate these
functions, and provides simple analytic continuation rules which
allow us to define the amplitudes throughout the complex plane.

\subsec{The 3-Point Function}

Using conformal invariance, we choose the three external vertex 
insertion points
at $(z_1,z_2,z_3) =(0,1,\infty)$. Thus, the electro-statics
problem of the previous section reduces to finding the
equilibrium configuration of a density of charges, in the
presence of 3 fixed charges, located at $0$, $1$ and
$\infty$. The general formalism of the previous section readily
provides the answers for the distribution of these charges.

We begin by obtaining the values for the parameters $R_i$ of
\rsol\ by solving just the asymptotic conditions, given in
\aeqs.  Using 
the definition $a=a_1+a_2$ of \polnorm, we obtain
\eqn\threea{
  R_1=-R_2=4(a+1) 
}
The electric field function $\omega (z)$ is readily given by
\omsol, and we find 
\eqn\threerho{
    \omega(z)
   =
    \half\left[{\al 1\over z}+{\al 2\over z-1}
    -(a+2){\sqrt{(z-x_1)(z-x_2)}\over z(z-1)}\right]
}
The branch points $x_1,~x_2$, which correspond to the endpoints
of the lineal distribution of Liouville charges, are given by an
algebraic function of the charges $a_i$ as follows
\eqn\wonetwo{
  x_{1,2}=
  {(\al1+2)a+2
  \pm 2\sqrt{(\al1+1)(\al2+1)(a+1)} \over (a+2)^2}
  }
The distribution of charges is located on a quadratic branch cut
curve $\c C$ between the points $x_1,~x_2$, and the lineal
density of charges is given by
 \eqn\threerho{   \rho(z)
    =
    {1\over2\pi}(a+2){\sqrt{(z-x_1)(x_2-z)}\over z(1-z)}
    }
where it is assumed that $z$ is restricted to the branch cut
curve $\c C$. 

When the charges $a_i$ are all real and positive, we have 
\eqn\xineq{
0< x_1 < x_2 < 1
}
and the branch cut curve $\c C$ naturally lies on the real axis,
between the points $x_1,~x_2$, as required by the reality
condition \reality. In fact, the endpoints $x_1,~x_2$ stay away
from the fixed charges at 0 and 1.

When the charges $a_i$ are extended into the complex plane, the
end-points $x_1,~x_2$ also move into the complex plane. In
general, the location of the branch cut curve $\c C$ is now not
naturally given by symmetry arguments alone.  Nonetheless, our
formalism is still well-defined and the reality requirement of
\reality\ determines the position of the branch cut curve $\c C$
in a unique way. To find it, we require that the indefinite
integral of $\rho(w)$ be real along the curve $\c C$
parametrized by some real parameter $t$, as given in \abelreal.
Unfortunately, this equation is not easy to solve in any
analytic way. 
However, we do not
actually need the precise location of these curves to compute
the final electro-static energy of the configuration.

To compute the integrals that yield the electro-static energy
$\c E_0$, in the correlation function \treefinal, we need to
evaluate integrals involving $dw~\rho(w)$, rational functions of
$w$ and logarithms. It is standard procedure to {\em uniformize}
the integral by making the following change of variables from
 $w$ to a variable $t$ (which is not necessarily real)
\eqn\changevar{
  t\equiv \sqrt{w-x_1\over x_2-w}
  }
Clearly, the integration range where $w$ goes from $x_1$ to
$x_2$ converts into an integration range where $t$ goes from $0$
to $\infty$. Thus, a definite integral in $w$ along the full
branch cut may be converted as follows
\eqn\rhoint{\eqalign{
    \int_{x_1}^{x_2}\!\!\!dw\,\rho(w)f(w)
         =  &{a+2\over\pi}
    \int _0^\infty\!\!\!dt\, 
    \biggl [{1\over t^2+1} -{x_1\over t^2+x_1/x_2}
\cr \qquad & -
    {1-x_1\over t^2+(1-x_1)/(1-x_2)}\biggr ]
    \,f\left({x_2t^2+x_1\over t^2+1}\right)\cr
    }
  }
 We find the following basic integral, valid for $x,~y$
 arbitrary complex variables, to be useful 
\eqn\intformulas{
    \int_0^\infty \!\!\!dt \,{1\over t^2+x}\ln(t^2+y)
   ={\pi\over \sqrt{x}}\ln(\sqrt{x}+\sqrt{y})   
  }
Combining all integrals, we obtain a closed expression for the
electric potential
\eqn\omegapot{\eqalign{
\Omega (z) 
= & 
  (a+2) \ln  (\sqrt{z-x_1} + \sqrt {z-x_2}) /2 \cr
& - 
  a_1 \ln { \sqrt{x_1}\sqrt{z-x_2} + \sqrt{x_2}\sqrt {z-x_1}
       \over \sqrt{x_1} + \sqrt{x_2}} \cr
& -
  a_2 \ln {\sqrt{1-x_1}\sqrt{z-x_2} + \sqrt{1-x_2}\sqrt {z-x_1} \over
            \sqrt{1-x_1} + \sqrt{1-x_2}} \cr }
}
Using the explicit expressions for $\Omega(w)$ in \sdef\ we obtain 
the expression for the three point function as
\eqn\check{\eqalign{
    \c E _0
    &=-2(\alpha s)^2\biggl[ 
    (\al1+1)^2\ln(\al1+1) - \al1^2\ln\al1
    +(\al2+1)^2\ln(\al2+1) - \al2^2\ln\al2
    \cr&\qquad
    +(a+1)^2\ln(a+1)
    -(a+2)^2\ln(a+2)
    \biggr]\cr
    &=2(\alpha s)^2\sum_{i=1}^3 \left[
      (\al i+ 1)^2\ln(\al i+ 1) - \al i^2\ln\al i
    \right]\cr
    }
  }
which agrees with the result obtained in \AD.
In obtaining the last equality, we used the sum of the charges
in \simplesum.
We note that the final formula is symmetric in the three
``charges'' $\al i$, as it should be.
\subsec{The 4--Point Function}
The electrostatics problem corresponding to the 4--point
function is to find the electrostatic energy of the equilibrium
configuration of the point charges $\al i\ (i=1,2,3,4)$ and the
Liouville charge density $\rho(z)$. 
The location of the point charges may be taken to be
$(z_1,z_2,z_3,z_4) = (z_1,0,1,\infty)$ using conformal 
invariance. 
The point $z_1$ and the Liouville charge density need to be
determined from the condition of electrostatic equilibrium.
Using the asymptotic conditions on the electric field function
$\omega(z)$ given in \aeqs, we may eliminate $R_1$ and $R_2$ in
terms of the $a_i$ and $R_3$. Making use of the definition
$a=a_1+a_2+a_3$ of \polnorm, we find
\eqn\twoa{
  R_2 = 4+4a+(z_1-1)R_1 
  \qquad \qquad
  R_3 = -4-4a-z_1R_1
   }
Unlike the case of the three point function, we can {\em not}
determine all the $R_i$'s solely from the asymptotics of
$\omega(z)$.  The density function $\rho(z)$ may be obtained as
\eqn\fourdensity{
  \rho(z)={1\over2\pi}{\sqrt{-Q_4(z)}\over z(z-1)(z-z_1)}
   }
where
\eqn\fourzero{
  \eqalign{
    Q_4(z) &=(a+2)^2\prod_{i=1}^4\left(z-x_k\right)
    \cr    &= 
    \bigl [a_1 z(z-1) + a_2(z-z_1)(z-1) 
    + a_3 (z-z_1)z\bigr ]^2 \cr
    & \quad +z(z-1)(z-z_1)\bigl [(4+4a) (z-z_1) 
    +R_1 z_1(1-z_1) \bigr ] \cr
    }
  }
and
\eqn\roneis{
  R_1 = -8\left[ {b_{12}\over z_1} + {b_{13}\over z_1-1}\right] 
  }

The locations of the zeros of $\rho(z)dz$, $\{x_k | k=1,2,3,4\}$
may be determined {\em algebraically} as a function of $z_1$ by
solving the fourth order polynomial equation $Q_4(z)=0$, albeit
cumbersome.  These zeros, together with the
reality condition \reality\ determine the support for the
Liouville charge density $\c C$.  

The quadratic branch cuts may be
chosen along the intervals $[x_1, x_2]$ and $[x_3 ,x_4]$; A- and
B-periods of the elliptic curve may be taken as the contours around
$[x_1, x_2]$ and the closed contour (along the first and second
sheets) $[x_2, x_3]$. 

The modulus $\tau$ of the curve is then determined from the periods,
which are given in turn by the elliptic integrals
\eqn\modulus{
  \tau = { \int _B dz/\sqrt { Q_4 (z)} \over \int _A dz /\sqrt
    {Q_4(z)}}
}
with the reality requirement on the A-period
\eqn\reala{
  \int _A dz /\sqrt {Q_4(z)} \ \ {\rm real}
}
An alternative determination of the modulus $\tau$ that does not
require explicit knowledge of the roots of $Q_4(z)$ proceeds from
the discriminant $\Delta$ of the curve $y^2 = Q_4(z)$. In general,
the discriminant is given in terms of the roots $x_i$ as follows
\eqn\disc{
\Delta = 16 \prod _{i<j\atop i,j=1}^4 (x_i - x_j )^2
}
The explicit expression for the zeros is however not needed, and
$\Delta$ may be expressed as a polynomial in the coefficients of
$Q_4(z)$. Equivalently, using a M\"obius transformation, one of the
zeros of $Q_4(z)$ may be moved to $\infty$ and the polynomial may be
put in Jacobi standard form
\eqn\jacobi{
\tilde Q_4 (w) = 4 w^3 -g_2 w -g_3
}
so that $\Delta = g_2 ^3 - 27 g_3 ^2$. 
The coefficients $g_2,g_3$ may be related to the coefficients of
the original polynomial $Q_4(z)$; let
\eqn\qqff{
  Q_4(z)\equiv \qq0 z^4+4\qq1 z^3+6\qq2 z^2 + 4 \qq3 z+\qq4
}
then
\eqn\gdeq{
  g_2=\qq0\qq4+3\qq2^2-4\qq1\qq3,\qquad
  g_3=\det\pmatrix{\qq0&\qq1&\qq2\cr\qq1&\qq2&\qq3\cr
    \qq2&\qq3&\qq4\cr}
}
It is straightforward but cumbersome to reexpress $g_2,g_3$ in
terms of physical quantities using \fourzero\ and we shall not
do so here.

The discriminant then determines the modular invariant function
$J$ as follows
\eqn\jfunction{
  J= {g_2 ^3 \over g_2 ^3 - 27 g_3 ^2}
}
and  $J$ determines the modulus $\tau$ in terms of a
hyperelliptic   expression
\eqn\hyperell{
  \tau = e^{2\pi i /3} 
  { F(J) - \mu e^{i\pi/3} J^{1 / 3} \tilde F (J)
    \over
    F(J) - \mu e^{-i\pi/3} J^{1 / 3} \tilde F (J)}
}
   where
\eqn\fdef{
  F(J) \equiv {}_2 F _1 ({1\over 12},{1\over 12}; 
  {2 \over 3};J),
  \qquad
  \tilde F(J)\equiv {}_2 F _1 ({5\over 12},{5\over 12}; {4 \over 3};J)}
and
\eqn\fdefthree{
   \mu \equiv (2-\sqrt{3}) {F(1) \over \tilde F(1)} 
        = (2 - \sqrt {3}) {\Gamma (11/12)^2 \Gamma (2/3) 
           \over \Gamma (7/12)^2 \Gamma (4/3)}}
While these expressions are not elementary, they are completely
explicit.

The density may be obtained in terms of geometric objects as in
\rhois 
\eqn\rhoone{
  dw\rho(w)=\sum_{i=1}^N \al i \tau_{z_i^+z_i^-}
  +\gamma_1\omega_1
}
where 
\eqn\gammaone{
  \gamma_1=2\int_{x_{1}}^{x_{2}}\!\!\!dw\,\rho (w)
  }
This determines $\gamma_1$ in terms of the external charges $\al 
i$'s and $R_1$.
Given the density $dw\rho$, it remains to integrate this function to
obtain the correlation function as in \ezerofinal\ --- numerically if
necessary. We shall not attempt to do this here, but consider a
simplified scattering configuration instead.

\subsec{Symmetric Scattering : A Solvable Case}
While the most general four point function does not seem to be
calculable analytically, we will treat an illustrative example
of a class of four point functions which is solvable
algebraically. It corresponds to the case where the external 
momentum distributions are symmetrical, and involve only
symmetric scattering.

We treat the case when the charge distribution of the
electrostatics problem corresponding to the 4--point function
has a $\IZ_2$ symmetry. For the class of 4--point functions
whose saddle points have this property, we may obtain the
Liouville charge density $\rho(z)$ and the four point function
algebraically. Without loss of generality, we may choose the
point charges $\{\al i|i=1,2,3,4, \ \al3=\al1\}$ to be at
$(z_1,z_2,z_3,z_4)=(-1,0,1,\infty)$.  The $\IZ_2$ symmetry
operation interchanges $z$ and $-z$.  This class of charge
configuration spans a subspace with complex codimension two of
the full configuration space.
The external momenta need to satisfy
\eqn\momentaeq{
  k_1k_2=k_2k_3
  } 
Since
  $\beta_1=\beta_3$, particles 1 and 3 should be regarded as both
incoming and \momentaeq\ corresponds to
  the condition of transverse scattering in the center of mass
  frame.  

{}From the asymptotics of $\omega(z)$, \aeqs, and the $\IZ_2$
symmetry we obtain $R_i$'s as
\eqn\asymeq{
  R_2=0,\qquad R_1=-R_3=2\left(a+1\right)
  }
Th is allows us to determine the zeros of the $\rho(z)$ and hence 
$\rho(z)$ itself as 
\eqn\rhosym{
  \rho(z)={1\over2\pi}(a+2)
    {\sqrt{(z^2-x_1^2)(x_2^2-z^2)}\over z(1-z^2)}
    }
where
\eqn\xsym{
  x_1^2,x_2^2={a(\al2+2)+2\pm
    \sqrt{2(\al1+1)(\al2+1)(a+1)}    \over(a+2)^2}
  }
The support for the density, $\c C$ is determined by the zeros
of $\rho(z)$ and the condition \abelreal.  In particular, for
any real values of $\al1,\al2$ (not necessarily positive) it may
be shown that $0\leq x_1,x_2\leq1$ so that the support for
$\rho$ is $\c C =[-x_2,-x_1]\cup [x_1,x_2]$.  We note that this
particular class of the 4--point function is identical to the
three point function except that $\al1$ and $z$ are ``doubled''
to $2\al1$ and $z^2$.  There are two further conditions that
need to be satisfied for consistency; namely $R_2$ defined by
the integral as in \summaryom\ needs to agree with $R_2$
obtained above from the asymptotics of $\omega(z)$ and the
integral of $\rho(w)$ over $\c C$ needs to be one.  A
straightforward analysis suffices to show that these identities
are indeed satisfied.

Once we have obtained the density $\rho(w)$, we may proceed to
compute the 4--point function using \sdef.
The computation may be carried out completely analytically as in 
the 3--point function case:
\eqn\fourpointsym{\eqalign{
    \c E_0
    &=
    -2(\alpha s)^2\biggl[ 
    (2\al1+1)^2\ln(2\al1+1) - 2\al1^2\ln(2\al1)+
    (\al2+1)^2\ln(\al2+1) - \al2^2\ln\al2
    \cr&\qquad
    +(a+1)^2\ln(a+1)   -(a+2)^2\ln(a+2)
    \biggr]
    \cr&=
    -2(\alpha s)^2\biggl[ 
    (2\al1+1)^2\ln(2\al1+1) - 2\al1^2\ln(2\al1)+
    \sum_{i=2,4}\left((\al i+1)^2\ln(\al i+1) - \al i^2\ln\al
      i\right)  
    \biggr]
    \cr
    }
  }
The above result may be naturally be continued to the complex
domain analogously to the three point function case.
The electrostatic energy behaves linearly as a function of the
Mandelstam variables $u_{ij}$, in the high energy limit, and thus
the 4-point amplitude for symmetrical scattering behaves
exponentially as a function of the Mandelstam variables. This
behavior is the one obtained for the high energy limit of critical
strings, yet the precise behavior of the two amplitudes is clearly
different. Also, even though exponential behavior was established
through explicit formulas only for the symmetric 4-point function,
it is clear that the expression for the general 4-point function
also exhibits exponential behavior of the amplitudes. We shall
expand further on this behavior in the next section.
\newsec{Discussion}
In the present paper, we have derived explicit formulas for the
high energy limit of non-critical string theory, on worldsheets
with the topology of the sphere. It is found that, for generic
values of the matter central charge $c$, the amplitudes behave
exponentially in the Mandelstam variables $u$, for large energy.
\eqn\exponential{
\langle \prod _{i=1} ^N {\cal V}_i \rangle \sim \kappa _1 e
^{\kappa _0 u}
}  
This behavior is analogous to that of critical strings at high
energies.  However, the precise argument $\kappa _0$ of the
exponential is different than in the critical case. In general, the
large energy limit is given in terms of hyper-elliptic integrals,
which, as we showed here, can be described concretely. Many very
interesting questions remain, which we shall  briefly address below.

\item{(1)} While the high energy asymptotic behavior of the
  non-critical amplitudes is generically exponential, it is in
  principle possible that the argument of the exponential $\kappa
  _0$, or that the prefactor in front of the exponential $\kappa _1$
  vanishes at certain special values of the central charge $c$. We
  have shown, by explicit calculation, that the argument of the
  exponential $\kappa _0$ does not vanish for the 4-point function
  of a symmetrical scattering process. By analyticity, it is then 
  expected that the argument will in fact be generically
  non-zero for {\bf any} value of the central charge.
  \nl\indent
  Cancellations of the prefactor $\kappa _1$ on the contrary are
  expected to occur at certain isolated values of the central charge
  $c$, such as at $c=1$, where the behavior of the amplitude is known
  to be polynomial in the Mandelstam variables. Within the limitations
  of the leading behavior of the high energy limit of the scattering
  amplitudes, this prefactor $\kappa _1$ is not accessible. Instead,
  its study would require going to next order and studying the small
  oscillation problem around the semi-classical dominant configuration.
  Zeros in $\kappa _1$ may then occur through the existence of
  zero modes of the small oscillation problem (which would be
  relatively easy  to see).  They could also arise through the
  vanishing of the regularized functional determinant of the small
  oscillation problem, such as could occur in $\zeta$-function
  regularization (which would be much more difficult to see).

\item{(2)} In this paper, we have calculated explicitly only in
  the case of tree level amplitudes. The higher loop case will
  be dealt with in a forthcoming publication. We shall just
  point out here that, in the case of higher loop amplitudes,
  the high energy limit of the amplitudes again corresponds to
  an electro-statics problem, this time on a higher genus
  Riemann surface. On physical grounds, it is easy to argue that
  the presence of the Liouville interaction will again produce
  an infinite number of charges, just as in the tree level case.
  The Liouville charges accumulate onto curve segments
  which produce quadratic branch cuts on the Riemann
  surface. This may be seen, for example, by using the method of
  images to represent the charges on higher genus surfaces, or
  may be worked out directly in terms of the prime form on the
  surface.  {}From these considerations, it is to be expected
  that the general behavior of the loop amplitudes is again
  exponential in the Mandelstam variables at high energy.

\item{(3)} Finally, the questions of practical importance, such
  as the implications of our results for the 3-D Ising model and
  off-shell string theory remain to be worked out in a concrete
  way. We expect that the high energy limit, derived here, can
  be used as the starting point for a systematic expansion to
  physical correlation functions. Clearly, much work remains to
  be done to achieve this goal, but we regard our results so far
  as a promising step in the right direction.

\nl\nl
\noindent{\bf Acknowledgments:}\par
We are happy to thank Yoshihisa Kitazawa and Norisuke Sakai for 
collaboration in the early stages of this research, and we are grateful
to Jean-Loup Gervais for discussions of his work on Liouville theory.
We have benefited from conversations with S.~Higuchi, S.~Nussinov,
D.H.~Phong and H.~Sonoda.
We thank the Aspen Center for Physics, and the CERN theory division,
for their hospitality while part of this work was being carried out.
We acknowledge support from a US/Japan exchange grant
INT-9315099 from the National Science Foundation and the Japan
Society for the Promotion of Science. E. D. was supported in part 
by the National Science Foundation, under grants PHY-92-18990 and 
PHY-95--31023, while K. A. was supported in part by the Ministry 
of Education, Science, Sports and Culture for Grant--in--Aid and Keio
University.
\appendix{A}{3-Point Function by Jacobi Polynomials}
For the case of the 3-point function ($N=3$), the
electro-statics problem may be solved explicitly for any finite
number $s$, and general complex valued charges. The method is a
generalization to complex charges of a technique presented in
\ref\SZE{G. Szeg\"o, {\sl Orthogonal polynomials}, American
  Mathematical Society, Providence, RI (1981)}. 
One introduces the polynomial $P_s(w)$ with roots $w_p$
\eqn\psdef{
  P_s(w)= \prod _{p=1} ^s (w-w_p)
}
We show that this polynomial satisfies the differential equation
for Jacobi polynomials
\eqn\jacob{
  (1-w^2) y'' + (\beta - \alpha -(\alpha +\beta +2)w )y' +
  s(s+\alpha +\beta +1)y=0
  }
in the following way. Since $P_s$ is of degree $s$, the
polynomial on the left hand side of \jacob\ obtained by setting
$y(w) = P_s (w)$, is at most of degree $s$. Thus, it suffices to
check that it vanishes on $s$ points, which we choose to be
$w_p$, $p=1,\cdots, s$. At these points, the third term in $y$
vanishes; working out the derivatives using \psdef, and then
using \jacob, we have
\eqn\jacobeq{
{P_s '' (w_p) \over P_s '(w_p) }
   = \sum _{q\not= p} ^s { 1 \over w_p - w_q}
   = {-1 -\alpha \over w_p -1} + { -1 -\beta \over w_p +1}
}
The above equation can be identified with the electro-statics
equilibrium equation for the $w_p$ for three external points
$(z_1, z_2, z_3)= (-1, 1, \infty)$ and charges related by
\eqn\charges{
    \alpha =  -1 +{a_1 s\over2},\qquad
    \beta  =  -1 + {a_2 s\over2} 
    }
Thus, the Liouville charges at $w_p$ are located at the zeros 
of the Jacobi
polynomials $P_s(w)=P_s^{(\alpha~ \beta)} (w)$. Jacobi
polynomials are perfectly well-defined for general complex
$\alpha $ and $\beta$. For $\alpha, ~ \beta$ real and $ > -1$,
the zeros are real and between $-1$ and $1$; this is in fact the
case originally considered in \SZE\ of all repulsive charges.
For general complex values of the charges, the zeros move into
the complex plane. For large $s$ however, the zeros always
accumulate on a curve of charge density, converging towards
$\rho (w)dw$ of \threerho.
\vfill\eject
\listrefs
\end